\documentclass[12pt]{article}
\usepackage{hyperref}
\usepackage{cite}
\usepackage{color}
\usepackage{graphicx}
\usepackage{amsmath}
\usepackage{amssymb}
\usepackage{xspace}

\makeatletter
\@addtoreset{equation}{section}

\makeatletter
\renewcommand\section{\@startsection {section}{1}{\z@}%
                                   {-3.5ex \@plus -1ex \@minus -.2ex}
                                   {2.3ex \@plus.2ex}%
                                   {\normalfont\large\bfseries}}
\renewcommand\subsection{\@startsection{subsection}{2}{\z@}%
                                     {-3.25ex\@plus -1ex \@minus -.2ex}%
                                     {1.5ex \@plus .2ex}%
                                     {\normalfont\bfseries}}

\def\baselinestretch{1.2}
\newcommand{\be}{\begin{equation}}
\newcommand{\ee}{\end{equation}}
\newcommand{\beq}{\begin{eqnarray}}
\newcommand{\eeq}{\end{eqnarray}}

\newcommand{\tr}{{\rm Tr}}
\newcommand{\gone}[1]{{}}

\newcommand{\N}{{\mathcal{N}}}

\begin{document}
\begin{titlepage}
\begin{flushright}
MAD-TH-10-02
\end{flushright}

\vfil

\begin{center}

{\bf \Large Branes and fluxes in special holonomy manifolds\\ and cascading field theories}

\vfil

Akikazu Hashimoto$^a$, Shinji Hirano$^b$, and Peter Ouyang$^a$

\vfil

$^a$ Department of Physics, University of Wisconsin, Madison, WI
53706, USA

$^b$ The Niels Bohr Institute, Blegdamsvej 17, DK-2100 Copenhagen,
Denmark

\vfil

\end{center}

\begin{abstract}
\noindent

We conduct a study of holographic RG flows whose UV is a theory in 2+1
dimensions decoupled from gravity, and the IR is the ${\cal N}=6,8$
superconformal fixed point of ABJM. The solutions we consider are
constructed by warping the M-theory background whose eight spatial
dimensions are manifolds of special holonomies $sp(1)\times sp(1)$ and
$spin(7)$. Our main example for the $spin(7)$ holonomy manifold is the
$A_8$ geometry originally constructed by Cvetic, Gibbons, Lu, and
Pope. On the gravity side, our constructions generalize the earlier
construction of RG flow where the UV was ${\cal N}=3$
Yang-Mills-Chern-Simons matter system and are simpler in a number of
ways. Through careful consideration of Page, Maxwell, and brane
charges, we identify the discrete and continuous parameters
characterizing each system.  We then determine the range of the
discrete data, corresponding to the flux/rank for which the
supersymmetry is unbroken, and estimate the dynamical supersymmetry
breaking scale as a function of these data.  We then point out the
similarity between the physics of supersymmetry breaking between our
system and the system considered by Maldacena and Nastase.  We also
describe the condition for unbroken supersymmetry on class of
construction based on a different class of $spin(7)$ manifolds known
as $B_8$ spaces whose IR is different from that of ABJM and exhibit
some interesting features.

\end{abstract}
\vspace{0.5in}

\end{titlepage}
\renewcommand{\baselinestretch}{1.05}  

\section{Introduction}
\label{sec1}

One of the most important longstanding problems of string theory has
been to understand the field theory living on a stack of M2-branes in
M-theory.  Using the AdS/CFT correspondence, it has been conjectured
that in an appropriate decoupling limit, the low-energy M2-brane
theory has a gravity dual description given by eleven-dimensional
supergravity on an $AdS_4 \times S^7$ background
\cite{Maldacena:1997re}.  Using the gravity dual picture, the
conjecture implies that the decoupled theory is a superconformal field
theory with ${\cal N}=8$ supersymmetry.  Moreover, this field theory
should arise as an infra-red fixed point of $U(N)$ supersymmetric
Yang-Mills theory in 2+1 dimensions \cite{Itzhaki:1998dd}. However, an
effective field theory description of the infrared fixed point theory
by itself has remained elusive.  One sign that the M2-brane theory
must be quite special is that the number of degrees of freedom scales
with the number $N$ of M2-branes in the stack as $N^{3/2}$.  The
description of a field theory which captures the $N^{3/2}$ scaling
remains a deep unsolved mystery.

\begin{figure}[t]
\centerline{\includegraphics[width=2.5in]{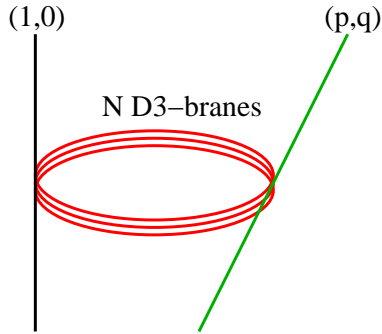}}
\caption{A configuration of D3, NS5, and $(p,q)$ 5-branes in type IIB
string theory. $N$ D3-branes wind around an $S^1$ of size $L$. An
NS5-brane and a $(p,q)$ 5-brane intersects the D3-brane at a localized
point along the $S^1$ but extends along the other 3 world volume
coordinates of the D3-branes.  Low energy effective theory of open
strings is a Yang-Mills/Chern-Simons/matter theory with gauge group
$U(N) \times U(N)$.
\label{figa}}
\end{figure}

An important step in addressing this problem was the construction of a
$2+1d$ Chern-Simons/matter theory with ${\cal N}=8$ supersymmetry by
Bagger, Lambert, and Gustavsson
\cite{Bagger:2007jr,Gustavsson:2007vu}.  Initially, the model of
Bagger, Lambert, and Gustavsson involved an exotic algebraic structure
known as a ``3-algebra'' in order to overcome some of the difficulties
encountered in \cite{Schwarz:2004yj}. It was later understood that
this 3-algebra structure can be mapped to the structure of a gauge
group $SU(2)\times SU(2)$ which has a product structure and is
therefore not simple \cite{VanRaamsdonk:2008ft}.  This construction
was further generalized by Aharony, Bergman, Jafferis, and Malcacena
(ABJM) in \cite{Aharony:2008ug} to a $U(N)_k \times U(N)_{-k}$ theory
which admits a construction based on Hanany-Witten-like setup
involving the D3-branes, NS5-branes, and $(p,q)$ five branes,
originally developed by \cite{Kitao:1998mf,Bergman:1999na} to describe
2+1 dimensional gauge theories with Chern-Simons terms. In this
construction, one of the world volume coordinates of the D3-brane is
compact, we take $(p,q) = (1,k)$, and the NS5 and the $(1,k)$ 5-brane
intersect the D3 at a point along this $S^1$, as shown in figure
\ref{figa}.  At low energies, the resulting theory is a $U(N)_k \times
U(N)_{-k}$ Chern-Simons theory where the subscript $k$ refers to the
(integer-valued) level of the Chern-Simons term, and a concrete
Lagrangian description is known.  Unfortunately, the decoupled theory
on M2 has $k=1$, which makes the theory strongly coupled. As a result,
this Lagrangian description is not so helpful in shedding new light on
the $N^{3/2}$ scaling unless one solves the theory exactly at strong
coupling. Nevertheless, the ABJM theory has many interesting features
in its own right.

One interesting fact about the ABJM theory is that it can be
constructed as the IR fixed point of an explicit renormalization
group flow.  Starting from the 5-brane construction in type IIB
string theory, one first takes the zero slope limit, which reduces
the string theory construction to a defect field theory on $R^{1,2}
\times S^1$. At energies below the scale set by the radius of the
$S^1$, the theory flows to a $U(N)_{k} \times U(N)_{-k}$
Yang-Mills/Chern-Simons/Matter theory in 2+1 dimensions.  As one
flows further to the IR, the Yang-Mills coupling becomes strong, and
we eventually flow to the superconformal fixed point of ABJM. This
RG flow can also be captured holographically
\cite{Hashimoto:2008iv}.

The ABJM theory may be further enriched by the inclusion of
``fractional'' D3-branes stretched between the NS5 and the $(1,k)$
5-brane. Aharony, Bergman, and Jafferis (ABJ) interpreted this
system as giving rise to a theory with $U(N)_{k} \times U(N+l)_{-k}$
where $l$ is the number of fractional D3-branes
\cite{Aharony:2008gk}.

In the far ultraviolet, the fractional branes can give rise to duality
cascades \cite{Aharony:2009fc} similar to those of Klebanov and
Strassler \cite{Klebanov:2000hb}.  There are, however, a number of
features which are special to these 2+1-dimensional field theories.
In particular, as one can see from the Hanany-Witten brane
construction, there is an $s$-rule constraint on the number of
fractional D3-branes that can leave some fraction of the supersymmetry
unbroken.  For the simple case of pure ${\cal N}=1$ $SU(N)_k$
supersymmetric Chern-Simons theory, there is indeed a known bound, $k
> N/2$, on the range of rank and level for which a supersymmetric
vacuum exists \cite{Witten:1999ds}. In light of this fact, it is
natural to wonder if the bound $k > N/2$ is somehow related to the
$s$-rule.

A more ambitious question concerns the fate of the dual supergravity
description of the NS5 $(1,k)$-brane system as the number of
fractional branes are pushed into the regime where the $s$-rule is
violated.  Presumably, in that regime, the dual supergravity solution
would capture the features of dynamical supersymmetry breaking.

In this article, we will study the gravity dual description of the
cascading renormalization group flow of the brane configuration of
figure \ref{figa} and closely related systems while focusing on the
fate of breaking supersymmetry by violating the $s$-rule.  Similar
issues were considered in a different construction by Maldacena and
Nastase \cite{Maldacena:2001pb}, and we will further elaborate on
similarities and differences between the two.

For technical reasons, the specific brane configuration of overlapping
NS5 and $(1,k)$ 5-branes turns out to unsuitable for our purposes. One
of the key ingredients in constructing the dual gravity description of
the cascading RG flow is a self-dual 4-form in an 8 dimensional
hyper-K\"ahler manifold known as the Lee-Weinberg-Yi (LWY) space
\cite{Lee:1996kz}. While there is a well known conjecture by Sen that
such a 4-form exists \cite{Sen:1994yi}, an explicit expression is not
known, preventing us from presenting completely explicit expressions
for the gravity dual.

However, there are a number of closely related RG flows which also
have the ABJM theory as the infrared fixed point, which we can study
quite explicitly.  One such example is a version of the LWY space with
$sp(1) \times sp(1)$ holonomy, and another is the asymptotically
locally conical geometry with $spin(7)$ holonomy discovered by
\cite{Cvetic:2001pga}.  We will elaborate on both of these examples in
the following sections.

Our main results can be summarized as follows. In the $sp(1) \times
sp(1)$ holonomy case we exhibit the supergravity solution in detail.
For each of the examples we study, we identify the $s$-rule bound
explicitly, and we identify the main physical mechanism responsible
for the breaking of supersymmetry in the dual gravity description.  In
all cases, the gravity dual interpretation of the supersymmetry
breaking is the dynamical generation of anti-brane charge in the
infrared.  Finally, although we do not attempt to find explicit SUSY
breaking solutions, we make some observations in the $spin(7)$ case
for how they might be constructed as the solution of a concrete system
of coupled ordinary differential equations.

We begin in Section \ref{sec2} by reviewing the brane construction and
the holographic dual of RG flow from the
Yang-Mills/Chern-Simons/matter theory to the ABJM fixed point. The
material contained in this section is mostly review, including the
status of the self-dual 4-forms in LWY space. In section \ref{sec3},
we describe the first alternate construction where ABJM fixed point is
embedded in a fixed point with ${\cal N}=4$ supersymmetry
\cite{Aharony:2008ug}.  Already with this construction, we can see the
signatures of breaking of supersymmetries when the $s$-rule is
violated.

In section \ref{sec4}, we describe another UV embedding of the ABJM fixed
point, this time involving a manifold of $spin(7)$ holonomy known as
the $A_8$ space, originally constructed by Cvetic, Gibbons, Lu, and
Pope \cite{Cvetic:2001pga}. This construction has the advantage that
the non-linear ansatz for the gravity solutions can be presented as a
function of a single coordinate, which allows us to explore the full
gravity ansatz in some detail.

In section \ref{sec5}, we comment on the interpretation of another
$spin(7)$ holonomy manifold known as the $B_8$ space
\cite{Cvetic:2001pga}. Here, the IR fixed point will not be of the
ABJM type, but we find a close connection between the supersymmetry
breaking of ABJM due to the violation of $s$-rule, and the dynamical
SUSY breaking of ${\cal N}=1$ Chern-Simons theory \cite{Witten:1999ds}
through the previous work of Gukov and Sparks \cite{Gukov:2001hf}. We
also comment on the interpretation of a specific deformation of $B_8$
space, also constructed by \cite{Cvetic:2001pga}, with flux which
breaks all supersymmetries, from a unified perspective. We then extend
these findings to a broader family of $B_8$-like spaces, known as
$B_{8+}$ and $B_{8-}$. As a bonus, we also find a new scaling limit of
$B_{8+}$, which we will call $B_{8\infty}$, which appears to have been
overlooked in the analysis of \cite{Cvetic:2001pga,Cvetic:2001zx}.

In the conclusion, we present some general discussions concerning the
expected low energy behavior of these models when the supersymmetry is
broken. We also provide an estimate of the supersymmetry breaking
scale for the theories near the threshold of breaking/restoring supersymmetry.

\section{Cascading solution with ${\cal N}=3$ Supersymmetry}
\label{sec2}

In this section, we review the construction which for an appropriate
set of parameters gives rise to a cascading field theory in 2+1
dimensions with ${\cal N}=3$ supersymmetry which flows in the IR the
ABJM whose supersymmetry is enhanced to ${\cal N}=6$ or ${\cal N}=8$.
Most of this section is a review of earlier work. The main ingredients
we review in this section are the details of the brane constructions,
the generalized $s$-rule criteria for supersymmetry, the dual
supergravity background, quantization of charges, and the status of
the self-dual 4-form.

\subsection{Brane construction of ${\cal N}=3$ cascading field theory}

Let us begin by reviewing the brane engineering construction of the
${\cal N}=3$ cascade. We begin in type IIB string theory with
D3-branes oriented along the directions 0126, where the $x^6$
direction is periodically identified. To this setup we add NS5-brane
extended along the directions 012345 and a $(1,k)$ 5-brane oriented
along\footnote{The notation $[3,7]_\theta$, for example, means that
the brane extends along a line in the $x^3$-$x^7$ plane at an angle
$\theta$ with respect to the $x^3$ axis.}
$012[3,7]_\theta[4,8]_\theta[5,9]_\theta$. To preserve at least six
supercharges, the angle $\theta$ must satisfy
\beq
\theta = {\rm arg}(\tau) -{\rm arg}(k+\tau)
\label{angle5brane}
\eeq
where $\tau$ is the IIB axiodilaton, $\tau = ie^{-\Phi} + C_0$. When
the background axion $C_0$ vanishes, $\tan\theta = 1/g_s k$.  This
gives rise to the configuration illustrated in figure \ref{figa}.

The configuration consisting of an overlapping NS5-brane and a $(p,q)$
5-brane separated in $x_6$ direction along with a D3 is extended was
considered extensively in
\cite{Kitao:1998mf,Bergman:1999na,Gukov:2002es}. The number of
unbroken supersymmetries in these configurations were also classified,
and we transcribe the result (originally reported in
\cite{Kitao:1998mf}) verbatim in table \ref{tablea}. The configuration
we consider corresponds to entry 4(iii) in this table. One difference
between the focus of \cite{Kitao:1998mf,Bergman:1999na,Gukov:2002es}
and our consideration is that we treat the $x_6$ direction to be
compact. This detail will turn out to have important consequences.

\begin{table}\begin{center}
\begin{tabular}{|c|c|c|c|c|}
\hline
Configuration & Angles & Condition & SUSY & second 5-brane \\
\cline{1-5}
1 & $\theta_4$ & $\theta_4 = 0$ & $\mathcal{N}=4$ & NS5 $(12345)$ \\
\cline{1-5}
2(i) & $\theta_2$, $\theta_3$ &
$\theta_2 = \theta_3$ &
$\mathcal{N}=2$ & NS5 $(123 [48]_{\theta_2} [59]_{\theta_3})$ \\
\cline{1-5}
2(ii) & $\theta_3$, $\theta_4$ &
$\theta_3 = \theta_4$ &
$\mathcal{N}=2$ & $(p,q)5$\ $(1234 [59]_{\theta_3})$ \\
\cline{1-5}
3(i) & $\theta_1$, $\theta_2$, $\theta_3$ &
$\theta_3 = \theta_1 + \theta_2$ &
$\mathcal{N}=1$ & NS5 $(12 [37]_{\theta_1} [48]_{\theta_2} [59]_{\theta_3})$ \\
\cline{1-5}
3(ii) & $\theta_2$, $\theta_3$, $\theta_4$ &
$\theta_3 = \theta_2 + \theta_4$ &
$\mathcal{N}=1$ & $(p,q)5$\ $(123 [48]_{\theta_2} [59]_{\theta_3})$ \\
\cline{1-5}
4(i) & $\theta_1$, $\theta_2$, $\theta_3$, $\theta_4$ &
$\theta_4 = \theta_1 + \theta_2 + \theta_3$ & $\mathcal{N}=1$ &
$(p,q)5$\ $(12 [37]_{\theta_1} [48]_{\theta_2} [59]_{\theta_3})$ \\
\cline{1-5}
4(ii) & $\theta_1$, $\theta_2$, $\theta_3$, $\theta_4$ &
$\theta_1 = - \theta_2$, $\theta_3 = \theta_4$ & $\mathcal{N}=2$ &
$(p,q)5$\ $(12 [37]_{\theta_1} [48]_{\theta_2} [59]_{\theta_3})$ \\
\cline{1-5}
4(iii) & $\theta_1$, $\theta_2$, $\theta_3$, $\theta_4$ &
$\theta_1 = \theta_2 = \theta_3 = \theta_4$ & $\mathcal{N}=3$ &
$(p,q)5$\ $(12 [37]_{\theta_1} [48]_{\theta_2} [59]_{\theta_3})$ \\
\hline
\end{tabular}\end{center}
\caption{Supersymmetric five-brane configurations in IIB theory.\label{tablea}}
\end{table}

If $N$ D3-branes are present, the field theory has gauge group $U(N)_k
\times U(N)_{-k}$ with Chern-Simons levels $k$ and $-k$ for the two
gauge group factors.  At low energies, the Chern-Simons terms give
masses to the vector multiplet.  Integrating out the vector multiplet
fields reduces the theory to a Chern-Simons matter theory with $\N=6$
superconformal symmetry (enhanced to $\N=8$ when $k=1$ or $k=2$.)

A natural generalization of the ABJM construction is to make the ranks
of the two gauge group factors unequal \cite{Aharony:2008gk}, so that
the gauge group is $U(N)_k\times U(N+l)_{-k}$.  This situation has a
simple description in terms of the IIB brane diagram.  One simply
takes $N$ D3-branes wrapped on the directions 0126 and $l$
``fractional'' D3-branes extended along 012 but with endpoints on the
NS5 and $(1,k)$ 5-brane in the $x^6$-direction.

In these brane constructions there is a moduli space associated with
the $N$ whole D3-branes, which may be moved freely in the $345789$
directions.  On the other hand, the $l$ fractional branes are not free
to move in the $345789$ directions because they end on the 5-branes,
and thus do not have corresponding moduli for generic values of the
angles $\theta_a$.

\subsection{Brane creation and the modified $s$-rule}

All of these considerations appear to give a consistent description of
the vacuum structure of the ABJM theories, but the ultraviolet
description of these theories is a bit more complex
\cite{Aharony:2009fc}. This is due to quantum mechanical effects in
the field theories \cite{Witten:1997sc}, and the Hanany-Witten brane
creation effect when the 5-branes cross each other
\cite{Hanany:1996ie}.

\begin{figure}
\centerline{\includegraphics{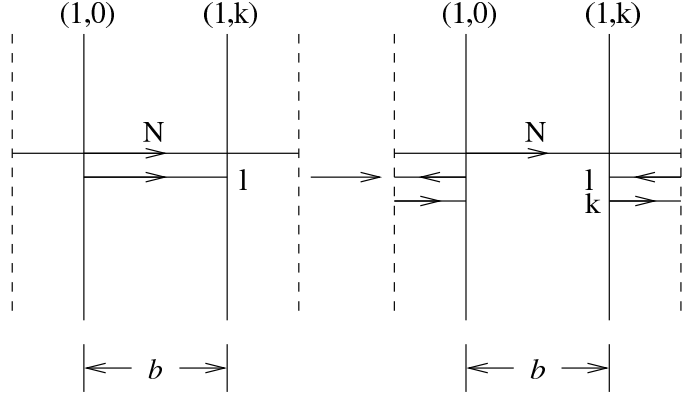}}
\caption{Illustration of brane creation effect when the $(1,k)$ 5-brane is moved around the periodic direction to the left.
\label{figb}}
\end{figure}

To set the stage, recall that when we move crossed 5-branes past each
other in type IIB, D3-branes extending between the 5-branes are
created \cite{Hanany:1996ie}.  Thus, for example, if we move the
$(1,k)$ 5-brane to the left in Figure \ref{figb}, $k$ D3-branes are
created and the resulting field theory has gauge group $U(N) \times
U(N-l+k)$.

In the far infrared, this gives a relation between two superconformal
Chern-Simons theories, as was studied in \cite{Aharony:2008gk};
specifically, one expects the two field theories to actually be fully
infrared-equivalent.  When $l<k$, this is a consistent procedure,
preserving $\N=6$ supersymmetry.

\begin{figure}
\centerline{\includegraphics{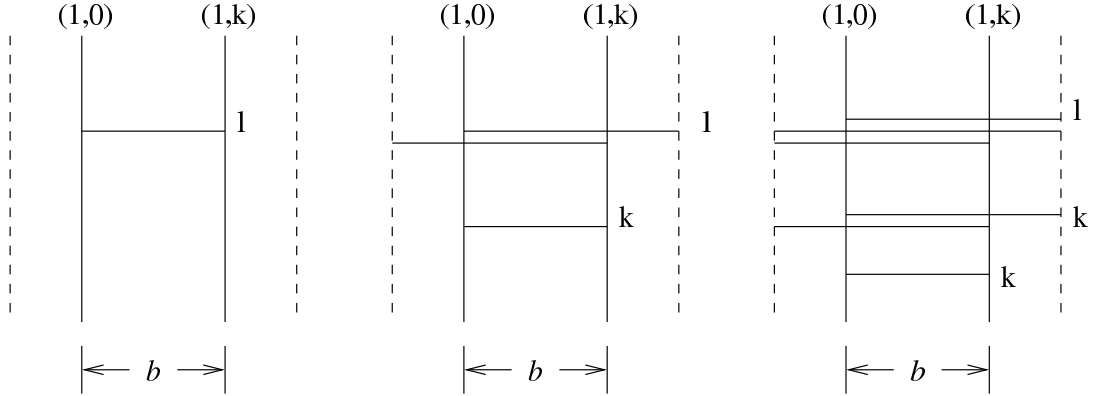}}
\caption{\label{figc}The brane configuration for the ${\cal N}=3$
theories, and its change upon ``sliding'' $b\to b+1 \to b+2$. In the
${\cal N}=3$ theories, fractional branes are {\it not} free to move in
the vertical direction.  We will nonetheless separate the branes in
the vertical directions to avoid cluttering the figure.  The $N$
integer branes, winding all the way around the periodic direction,
have also been suppressed in the figure.}
\end{figure}

Now, what happens if instead of moving the $(1,k)$ 5-brane to the
left, we move it to the right, around the $x^6$ circle?  Once again,
we expect the brane creation process to take place, as shown in Figure
\ref{figc}.  We see that after performing this transition, on one
interval between the two 5-branes, there are $N+2l+k$ 3-branes, while
there are $N+l$ 3-branes in the other interval.  In the UV, it is
tempting to identify the associated field theory with an $\N=3$ YM-CS
theory with gauge group $U(N+2l+k) \times U(N+l)$.  Continuing this
procedure $n$ times, one finds that the resulting brane configuration
has $N+nl + \frac12 n(n-1)k$ units of whole D3-brane charge, along
with $l+nk$ units of fractional brane charge, leading to a natural
identification with a YM-CS theory with gauge group $U(N+nl + \frac12
n(n-1)k)_k\times U(N+(n+1)l + \frac12 n(n+1)k)_{-k}$ in the UV.

Naively, the brane configurations that we obtain by moving the $(1,k)$
brane to the right appear to violate the $s$-rule, which says that
there can be at most $k$ D3-branes stretched between an NS5-brane and
a $(1,k)$ 5-brane \cite{Kitao:1998mf,Bergman:1999na,Hanany:1996ie}.
However, because our branes live on a circle, there is a subtlety in
interpreting the $s$-rule. In the case that we have a fractional brane
together with a regular brane, we could interpret this either as one
D3-brane stretched directly along the segment from the NS5-brane to
the $(1,k)$ 5-brane, and another D3-brane wrapping the circle, or as a
single D3-brane that winds around the circle more than once. In other
words, in the covering space of the circle, we can have D3-branes that
stretch between the NS5-brane and different images of the $(1,k)$
5-brane \cite{Dasgupta:1999wx}, and the ``modified $s$-rule'' just
tells us that there can be at most $k$ D3-branes stretched between the
NS5-brane and a specific image of the $(1,k)$ 5-brane. This condition
can be summarized, simply, by the condition
\be N > {l(l-k) \over 2k} \label{srule} \ee
which is invariant under shifts of $N$ and $l$ by $n$, and reduces
to $k > l$ for $N=0$. This modified $s$-rule is necessary for the
low-energy physics to be independent of the $x^6$ positions of the
5-branes, even in highly supersymmetric configurations such as those
of the previous section.\footnote{Similar modifications to the
conditions for supersymmetry occurs when one adds additional matter
to the theory \cite{Giveon:2008zn}. There are also evidence for a
rich phase structure as the parameter of these theories are varied
\cite{Giveon:2007fk}.}

The picture of a modified $s$-rule is consistent with the
interpretation of \cite{Bachas:1997sc,Bachas:1997kn}, who found that
in a dual frame (with F-strings stretched between D0 and D8 branes)
the $s$-rule is a manifestation of the Pauli exclusion principle.  The
fermionic modes living on the branes which wrap all the way around the
$x^6$ circle $m$ times are distinguishable from those that are on the
branes that wrap around the circle $m' \neq m$ times, so it is
possible for all of them to be in their ground state.

Note that this interpretation requires that the branes are connected
in a particular way, so it does not correspond to an infrared
statement about CS theories.  Indeed, the classical moduli space of
the $U(N+nl + \frac12 n(n-1)k)_k\times U(N+(n+1)l + \frac12
n(n+1)k)_{-k}$ YM-CS theory must receive corrections.  Naively, there
is a moduli space of dimension $8(N+nl + \frac12 n(n-1)k)$
corresponding to the motions of regular branes, but this cannot be the
case, because at generic points in moduli space, the resulting IR
theory would be a CS theory with gauge group $U(l + k)_{-k}$ which
does not have a supersymmetric vacuum. Instead, it was conjectured in
\cite{Aharony:2009fc} that the true moduli space receives quantum
corrections and is only $8N$-dimensional.

Given these subtleties in interpreting the brane diagram, one might
worry that the brane configurations obtained here do not make sense.
One of the results of this paper is the construction of an associated
gravity dual of these brane configurations that violate the ``naive''
$s$-rule but satisfy the ``modified'' $s$-rule, giving us confidence
that the picture presented above is in fact consistent.

The process of brane creation in the UV seems quite similar to the
duality cascades in the Klebanov-Strassler system
\cite{Klebanov:2000hb}.  Given some initial UV YM-CS theory, one
expects it to flow to a superconformal CS theory with a gauge group
with reduced ranks; in this regard the KS system is quite different,
in that its IR fixed point is confining.  Another important difference
is that in our case the cascade can terminate in the UV rather than
continuing indefinitely as in KS; this fact arises essentially because
our field theories are 2+1-dimensional and therefore asymptotically
free.  Finally, and perhaps most importantly, the YM-CS theory (and
its closely related cousins) have an $s$-rule bound on the
preservation of supersymmetry which has no analogue in KS.

\subsection{Gravity dual of the ${\cal N}=3$ cascade}

In this subsection, we will briefly review the dual gravity
description of the brane construction illustrated in figure
\ref{figa}. The supergravity background for intersecting brane
configurations is, in general, very difficult to find. For the
specific intersection of NS and $(p,q)$ 5-brane illustrated in figure
\ref{figa}, however, there is a well known solution in type IIA
obtained by T-dualizing along the $x_6$ direction, mapping NS5 and the
$(p,q)$ 5-branes to a pair of overlapping KK 5-branes
\cite{Gauntlett:1997pk}.  When this type IIA background is lifted to
M-theory, it takes the form
\be R^{1,2} \times {\cal M}_8 \ee
where ${\cal M}_8$ is a Ricci-flat hyper-K\"ahler manifold equivalent
to the metric on the hyper-K\"ahler moduli-space of dyons and is also
known as the Lee-Weinberg-Yi metric \cite{Lee:1996kz}. This space is a
$T^2$ fibration, and has a metric of the form
\beq
ds_{8}^2 = V_{ij} d\vec y_i d\vec y_j + (V^{-1})^{ij} R_i R_j (d
\varphi_i+ A_i) (d
\varphi_j + A_j),
\label{lwy}
\eeq
where
\beq
V_{ij} = V_{ij}^{\infty} + {1 \over 2} {R_i p_i R_j p_j \over |R_1
p_1
\vec y_1+ R_2 p_2
\vec y_2| }
+ {1 \over 2} {R_i \tilde p_i R_j \tilde p_j \over |R_1 \tilde p_1
\vec y_1+ R_2 \tilde p_2
\vec y_2| } \ ,
\label{defVij}
\eeq
$(i,j=1,2)$, $\vec{y}_1, \vec{y}_2$ are two 3-vectors, and
$\varphi_i
\equiv \varphi_i + 2\pi$.
generalizing the familiar 4-dimensional Taub-NUT metric. $R_1$ and
$R_2$ are the radius of the cycles parameterized by $\varphi_1$ and
$\varphi_2$, respectively. The pairs of integers $(p_1, p_2)$ and
$(\tilde p_1, \tilde p_2)$ encode the NS5 and D5 charges of the
5-branes in the type IIB description.  To obtain a geometry with
$(1,0)$ and $(1,k)$ 5-branes, we choose
\beq
(p_1,p_2) = (1,0), \qquad (\tilde p_1, \tilde p_2) = (1,k) \ . 
\eeq
In this case, the LWY geometry approaches $R^8/Z_k$ near the core.

One can add D3-branes winding all the way across the $x_6$ direction
in the original type IIB description, which corresponds to adding an
M2-brane in the M-theory description. If we take the action of
eleven-dimensional supergravity in the standard form\footnote{We will
follow the conventions of appendix B of \cite{Aharony:2008ug}.}
\beq
 S_{11} = {1 \over 2 \kappa_{11}^2} \int d^{11}x\,  \sqrt{-g} \left(R - {1 \over 2}
|G_4|^2 \right) - {1 \over  2 \kappa_{11}^2} \int {1 \over 6}   C_3
\wedge G_4 \wedge G_4,
\label{mlag}
\eeq
the effect of adding M2-branes can be captured by an ansatz of the form
\beq
ds^2 &=& H^{-2/3} (-dt^2+dx_1^2+dx_2^2) + H^{1/3} ds_{8}^2, \\
G_4 &=& dC_3 = dt \wedge dx_1 \wedge dx_2 \wedge d H^{-1},
\label{ansatz1}
\eeq
where the warp factor $H$ is a harmonic function on LWY space
\cite{Hashimoto:2008iv}.

In taking the zero-slope limit of this background, we scale
\be R_1 = 2 \pi {\alpha'} L, \qquad R_2 = g_s l_s = g_{YM2}^2 \alpha' \ee
keeping $L$ and $g_{YM2}$ fixed\footnote{\label{fn4}Here, $g_{YM2}^2$ refers to
an overall scale implied by the Yang-Mills coupling, and not the
Yang-Mills coupling $g_{YM2a}^2$ and $g_{YM2b}^2$ of $U(N_a) \times
U(N_b)$ gauge group, as we will elaborate further below.}, so that
both $R_1$ and $R_2$ scales as $\alpha'$. This will keep the period
$L$ of $x_6$ and the gauge coupling $g_{YM3}^2 = g_{YM2}^2 L$ in the
IIB frame finite. With this scaling, the full structure of LWY space
survives the zero slope limit \cite{Hashimoto:2008iv}, closely
resembling the similar analysis involving the case of Taub-NUT
geometry \cite{Cherkis:2002ir}. The finiteness of $L$ implies that
this decoupling limit retains the dynamics of a 3+1-dimensional defect
field theory.

The fact that the background described above is a T-dual along the
$x_6$ direction does imply that the structures localized along the
$x_6$ direction, such as the position of 5-brane impurities, are
obscure in this description (See \cite{Tong:2002rq} for an interesting
account of a closely related issue.) One critical piece of information
encoded in the $x_6$ coordinate is the distance $b$ separating the
NS5-brane from the $(p,q)$ 5-brane. Using the convention where $b$ is
the fraction of the period $L$ of the compact $x_6$ coordinate, the
strength of the gauge coupling in 2+1 dimensions takes the form
\be {1 \over g_{YM2a}^2} =  {b \over g_{YM2}^2}, \qquad
 {1 \over g_{YM2b}^2} =  {1-b \over g_{YM2}^2}\ . \ee
As is often the case with holographic descriptions of quiver gauge
theories, this data is encoded in the NSNS 2-form through a 2-cycle in
the geometry seen from the type IIA perspective.  The 2-cycle in
question arises from the $CP^1$ homology cycle inside the $CP^3$ base
of the LWY space reduced to type IIA on $\varphi_2$.

Generalization to the case where there are $l$ fractional D3-branes
corresponds to turning on a self-dual 4-form in ${\cal M}_8$ so that
the M-theory ansatz becomes
\beq
ds^2 &=& H^{-2/3} (-dt^2+dx_1^2+dx_2^2) + H^{1/3} ds_{8}^2, \\
G_4 &=& dC_3 = dt \wedge dx_1 \wedge dx_2 \wedge d H^{-1} + G_4^{SD}\ .
\label{ansatz2}
\eeq
Provided that $G_4$ is a self-dual 4-form in ${\cal M}_8$, this ansatz
will solve the equation of motion of 11 dimensional supergravity.

\subsection{Quantization of charges}
\label{sec24}

An important ingredient in interpreting the holographic duality of
field theories and gravity background is the quantization of discrete
field theory data on the gravity side. On gravity side, this arises
from flux quantization. In a background with non-trivial fluxes, care
is needed in identifying the appropriate fluxes for which to impose a
quantization condition. That this can be subtle in a theory of gravity
which includes a Chern-Simons terms was highlighted in an important
paper by Marolf \cite{Marolf:2000cb}. In particular, there are three
independent notions of charges, Page, Maxwell, and brane charges,
which can take on distinct values in a presence of non-vanishing
fluxes.  It is the Page charge,\footnote{We follow the convention in
appendix B of \cite{Aharony:2009fc}.}
\beq k &=& {1 \over 2 \pi l_s g_s} \int_{CP^1} F_2, \\
l -{k \over 2} &=& {1 \over (2 \pi l_s)^3 g_s}   \int_{CP^2} (- \tilde F_4)- B_2 \wedge F_2, \label{D4charge}\\
N &=& {1 \over (2 \pi l_s)^5 g_s} \int_{CP^3} * \tilde F_4 - B_2
\wedge (-\tilde F_4) + {1 \over 2} B_2 \wedge B_2 \wedge F_2 \ ,
\eeq
which satisfies a Gauss' law and admits an integer quantization
condition.
The $CP^1$, $CP^2$, and $CP^3$ refer to the 2, 4, and 6 cycles of the
$CP^3$ base of the type IIA geometry.  The extra contribution from
$k/2$ in (\ref{D4charge}) is due to the Freed-Witten anomaly
\cite{Freed:1999vc} whose role in this context was elaborated
extensively in \cite{Aharony:2009fc}. These and
\be b_\infty = {1 \over 4 \pi \alpha'}\int_{CP^1} B \ee
in the large radius limit encode all of the UV parameters of this
construction.

The D4 Page charge depends both on the gauge-invariant four-form flux
and the NS-NS two-form potential; in M-theory, this corresponds to a
dependence both on the four-form flux and on a pure-gauge component of
the three-form potential.  Specifically, the three form potential in
M-theory has the form
\be C_3 = m B_{(3)} + \alpha d \sigma \wedge d \varphi \label{c3}\ee
where
\be \sigma = d \varphi + {\cal A} \ee
is the one-form which is associated with the M-theory circle,
\be d \varphi = {2 \over k R_{11}} dx_{11} \ . \ee
The term proportional to $m$ is the one which gives rise to a
non-trivial four form field strength
\be G_4 = d B_{(3)} \ee
whereas the $\alpha$ term is exact (despite being pure gauge, it is
necessary to keep this term when the background geometry admits
discrete homology cycles.)

To convert to the Type IIA language, we adopt the convention that
reduction from M-theory to IIA is achieved by
\be C_3 = A_3 - B_2 \wedge dx_{11} \ . \ee
Then we will find that the Page flux
\be (2 \pi l_s)^3 g_s l = \int_{CP^2} - \tilde F_4 - B_2 \wedge F_2
= \int_{CP^2} - d (A_3 + B \wedge A_1)
= - \alpha \int_{CP^2} {2 \over k R} d \sigma \wedge F_2 \label{pageflux}
\ee
perhaps surprisingly is independent of $m$.  Rather, $m$ is associated
with the parameter $b_{\infty}$.

Once these UV parameters are fixed, one can imagine taking the limit
$L \rightarrow 0$, or equivalently, $R_1 \rightarrow \infty$, which
decouples the degrees of freedom corresponding to momentum modes along
the $x_6$ directions in the IIB picture (or winding modes in the IIA
picture), giving rise to a $2+1d$ YM-CS-Matter theory as the weakly
coupled UV fixed point.

\subsection{Self-dual 4-form in LWY geometry}

In earlier subsections, we reviewed most of the general features which
go into the construction of the ${\cal N}=3$ cascading field theory
and its supergravity dual.  On the gravity side, the primary such
feature is the presence of self-dual 4-form flux in the LWY geometry
${\cal M}_8$.  We will conclude our discussion of the ${\cal N}=3$
cascade by reviewing the status of this 4-form.

Finding the supersymmetric 4-form is in principle just a problem in
differential geometry.  The metric of the LWY manifold is known, so
one strategy for solving the problem is as follows.  First, one simply
writes out the general 4-form which is self-dual, tri-primitive, and
$(2,2)$.  These constraints are algebraic, so the 4-form is determined
up to a number of functions which depend on three variables
(consistent with the isometries of the LWY space.)  Then one imposes
the Bianchi identity, which gives a system of coupled partial
differential equations for these functions.  However, the large number
of functions makes the task of finding a solution of this system of
equations quite difficult.

The existence of such a 4-form is related to the spectrum of bound
states of monopoles and dyons and has long been conjectured to exist
by Sen based on consideration of S-duality \cite{Sen:1994yi}.  There
exists one reference in the literature \cite{Gibbons:1996wc} where
this very 4-form is claimed to be constructed. Our analysis of the
4-form in \cite{Gibbons:1996wc}, however, appears to suggest that this
form is not self dual as claimed.

While numerous qualitative conclusions can be inferred by making mild
assumptions concerning the nature of this 4-form, not having its
explicit form at hand is a significant disadvantage in exploring the
cascading theory in detail.  Fortunately, there are several close
relatives of the LWY geometry for which the self-dual 4-form can be
constructed explicitly. In the following sections, we will describe
two such constructions: the cascade with ${\cal N}=4$ supersymmetry on
an $sp(1)\times sp(1)$ holonomy manifold and the cascade with ${\cal
N}=1$ supersymmetry on $spin(7)$ holonomy manifolds.  As an
application of this construction, we will see an emergent pattern in
the way that breaking of supersymmetry is manifested as $N$, $l$, $k$,
and $b_\infty$ are tuned to the regime where supersymmetry is expected
to break.

\section{Cascade With $\mathcal{N}=4$ Supersymmetry}
\label{sec3}

The story presented up to this point has been in terms of an $\N=3$
brane configuration that is naturally related to a precisely
formulated YM-CS theory.  Unfortunately, the study of the
gravitational dual of this YM-CS theory proves to be extremely complex
(see \cite{Hashimoto:2008iv} for an attempt.)  We would like to be
able to study a related system instead with a bit more symmetry so
that the supergravity analysis will be more tractable.

Fortunately, it was pointed out in \cite{Aharony:2008ug} that there is
actually a different field theory preserving at least $\N=4$
supersymmetry at all scales which flows to the $\N=6$ ABJM
superconformal field theory in the infrared.  This theory is
constructed in terms of a IIB brane configuration by starting with the
same constituent branes as in the $\N=3$ case -- D3-branes on 0126, an
NS5-brane on 012345, and a $(1,k)$ 5-brane oriented on
$012[3,7]_\theta[4,8]_\theta[5,9]_\theta$ with the angle $\theta$
determined by eq. (\ref{angle5brane}).  To preserve $\N=4$
supersymmetry, we simply need to tune the Ramond-Ramond axion $C_0$ to
set $\theta=\pi/2$.

This $\N=4$ system preserves the features of the $\N=3$ brane
construction that we wanted to study.  The brane creation process
occurs in exactly the same way, so the duality cascade phenomenon
should be common to both systems.  The infrared in both cases is the
ABJM/ABJ superconformal field theory.  And the theories have the same
moduli spaces corresponding to mobile ``whole'' D3-branes and pinned
fractional branes.

One disadvantage of using this $\N=4$ system compared to the $\N=3$
system is that the precise nature of the field theory is somewhat
obscure, for a reason pointed out in \cite{Aharony:2008ug}.  In the
$\N=3$ IIB brane construction, the gauge theory was a Yang-Mills
theory with product gauge group, Chern-Simons terms for the two gauge
groups, and some additional matter fields.  In the $\N=4$ case, a
similar interpretation is not possible because YM-CS theories with
greater than $\N=3$ supersymmetry generically do not exist
\cite{Kao:1992ig,Kao:1993gs}.\footnote{The point is that the CS term makes the
gauge field massive, and a massive vector multiplet with $\N=4$
supersymmetry in three dimensions contains fields with spins 1, 1/2,
0, $-1/2$, and $-1$.  In a YM-CS theory there is no candidate for
the spin $-1$ state so it is not usually possible to realize the
$\N=4$ supersymmetry (there are exceptions for Abelian gauge groups,
and of course there can be more supersymmetry for CS theories
without YM kinetic terms.)}

So whatever the field theory is, it cannot be a weakly coupled
Yang-Mills-Chern-Simons theory.  One way out was proposed already in
\cite{Aharony:2008ug} -- the brane construction is necessarily in a
background with large dilaton so the field theory is strongly coupled
and need not have a straightforward Lagrangian interpretation.
Understanding this highly supersymmetric field theory seems to be an
interesting open problem.

The advantage of the ${\cal N}=4$ construction, however, is the fact
that the dual gravity description is particularly simple. Start, as
before, with an eight dimensional manifold ${\cal M}_8$ of LWY type
(\ref{lwy})--(\ref{defVij}) where we take the charges of the 5-branes
to be
\beq (p_1,p_2) = (1,0), \qquad (\tilde p_1, \tilde p_2) = (1,k) \ . \eeq

In order to preserve $\N=4$ supersymmetry we need to make the choice
\beq
V^{\infty}=
\left(\begin{array}{cc}
1+\frac{R_1^2}{(kR_2)^2} & \frac{R_1}{kR_2} \\
\frac{R_1}{kR_2}  & 1
\end{array}
\right)
\eeq
(With the choice $V^{\infty}_{ij}=\delta_{ij}$ we would have obtained
the LWY geometry corresponding to the $\N=3$ YM-CS theory.)  By making
a change of variables
\beq
&& \vec{w}_1= R_1\vec{y}_1, \qquad \vec{w}_2= kR_2\vec{y}_2 +
R_1\vec{y}_1\label{wvars} \\
&& \varphi_1'= \varphi_1-\varphi_2/k, \qquad \varphi_2' =
\varphi_2/k
\label{phiprime}\\
&& A_1' = A_1-A_2/k,\qquad A_2' = A_2/k\ , 
\label{Aprime}
\eeq
it becomes clear that the geometry is simply the direct product of
two Taub-NUT manifolds, quotiented by a $Z_k$ orbifold. Explicitly,
we have
\beq
ds_{8}^2 = U_{ij} d\vec w_i d\vec w_j + (U^{-1})^{ij}(d
\varphi_i'+ A_i') (d
\varphi_j' + A_j'),
\label{tntn}
\eeq
with
\beq
U=
\left(\begin{array}{cc}
\frac{1}{R_1^2}+\frac{1}{2w_1} & 0\\
0  & \frac{1}{(k R_2)^2}+\frac{1}{2w_2}
\end{array}
\right)\equiv
\left(\begin{array}{cc}
U_1 & 0\\
0  & U_2
\end{array}
\right) \ . 
\eeq
As we see, the matrix $U$ is diagonal; it was pointed out in
\cite{Gauntlett:1997pk} that this special case of the LWY metric
preserves eight supercharges.

With respect to the spherical coordinates defined in relation to
$\vec{w}_1, \vec{w}_2$, the vector potentials $A_i'$ satisfy
\beq
dA_1' &=& \frac12 d\theta_1 \wedge \sin\theta_1 d\phi_1\\
dA_2' &=& \frac{1}{2} d\theta_2 \wedge \sin\theta_2 d\phi_2.
\eeq
The coordinates $\theta_i,\phi_i$ are defined over the standard ranges
$0\le\theta_i<\pi$ and $0\le \phi_i < 2\pi$ while the $U(1)$ fiber
coordinates range over $0\le \varphi_1< 2\pi$, $0\le \varphi_2<
2\pi/k$.  The identification of the $\varphi_i$ coordinates makes it
clear that the geometry is the $Z_k$ orbifold of Taub-NUT times
Taub-NUT, and the small-radius geometry is the orbifold $C^4/Z_k$
where the orbifold acts on each of the $C^1$ factors in the same
way. This geometry has the isometry group $SO(4) \times U(1) \times
U(1)$.  We will see that this system is symmetric enough for us to
find an analytic solution of 11-dimensional supergravity.

\subsection{Supersymmetry preserving  4-form flux in $TN\times TN/Z_k$}

The main advantage of the ${\cal N}=4$ construction is the relative
ease with which we can add the background four-form in the internal
eight-dimensional manifold.  We can preserve supersymmetry if the
flux obeys some simple properties: it must be self-dual and
primitive, with index structure (2,2) \cite{Becker:2001pm}.  To have
a well-defined supergravity solution it is also necessary for the
four-form flux to be $L^2$ normalizable, and to satisfy the Bianchi
identities of eleven-dimensional supergravity, the four-form must be
closed.

There is a natural candidate for the four-form which satisfies these
properties.  Recall that each Taub-NUT space in our transverse
8-manifold has a natural (1,1) anti-self-dual 2-form.  In the
coordinates of the previous section, the anti-self-dual two-forms
are
\beq
P^i =\frac{1}{U_i^2}
\left[\frac{dw_i}{w_i^2}\wedge(d\varphi_i'+A_i')
 + U_i d\theta_i\wedge \sin\theta_i d\phi_i\right]
 = d \left[ \frac{2}{U_i} \left(d\varphi_i'+A_i'\right) \right].
\eeq
Note that the two-form $P_i$ vanishes as $w_i\rightarrow 0$.

The wedge product of these two (1,1) forms,
\beq
\Omega_4 \equiv  P^1 \wedge P^2,
\eeq
satisfies all the necessary properties for supersymmetric four-form
flux in the full 8-dimensional manifold\footnote{This proposal has
actually appeared previously in the literature in the more general
context of a self-dual 4-form for the $\N=3$ LWY geometry
\cite{Gibbons:1996wc}.  In the $\N=4$ case the expression of
\cite{Gibbons:1996wc} reduces to our four-form. In the $\N=3$ case
the analogous ansatz fails to be co-closed.}. It also preserves all
the isometries of the geometry. The four-form $\Omega_4$ inherits
the properties of closure and $L^2$ normalizability from the parent
Taub-NUT spaces, as well as the (2,2) index structure. Up to an
overall orientation convention, $\Omega_4$ is self-dual.

The only property left to check is primitivity.  This is most
straightforward in a vielbein basis:
\beq
\vec{E}_1 &=& U_{1}^{1/2}
d\vec{w}_1\\
E^1_0 &=& U_{1}^{-1/2}(d\varphi_1' + A_1')\\
\vec{E}_2 &=&U_{2}^{1/2}
d\vec{w}_2\\
E^2_0 &=& U_{2}^{-1/2} (d\varphi_2' + A_2').
\label{neq4vielbeins}
\eeq
The three K\"ahler forms are given by
\beq
J_a = J^1_a + J^2_a
\eeq
with
\beq
J^i_a = E^i_0 \wedge E^i_a +\frac12 \epsilon_{abc} E^i_b \wedge
E^i_c
\eeq
and the indices $a,b,\ldots =1,2,3$.  In terms of the vielbeins we
have
\beq
P^i = \frac{1}{U_i^2}\frac{w^a_i}{w_i} \left( E^i_0 \wedge E^i_a
-\frac12 \epsilon_{abc} E^i_b \wedge E^i_c\right).
\eeq
In this form one can easily check that $P^i \wedge J^i_a=0$, so that
the four-form $\Omega_4$ is primitive with respect to all three
K\"ahler forms.

Thus the cascading solution has a four-form flux given by
\beq
G_4 &=& dC_3 = dt \wedge dx_1 \wedge dx_2 \wedge d H^{-1} +
\frac{\pi l_p^3}{2}{q \over R_1^2 (kR_2)^2}
\Omega_4,
\label{ansatz3}
\eeq
where the normalization for $\Omega_4$ has been chosen for the
following reason.  In the asymptotic limit where both the $w_i
\rightarrow \infty$,
\beq
\Omega_4 \rightarrow R_1^2(kR_2)^2 d\theta_1 \wedge \sin\theta_1 d\phi_1
\wedge d\theta_2 \wedge \sin\theta_2 d\phi_2
\eeq
and integrating over the natural $S^2 \times S^2$ cycle, the total
number of units of M5-brane charge\footnote{This quantity
corresponds to the Maxwell charge in the IIA description, which we
will describe more fully in section \ref{sec32}.}  is given by
\beq
q= \frac{1}{(2\pi l_p)^3} \int_{S^2\times S^2} (-G_4) \ .
\eeq

\subsection{Fluxes and Quantization}
\label{sec32}

To interpret this supergravity solution, we need to match the
parameters characterizing it to the field theory.  Some of these
parameters must be quantized (both from the point of view of the
string theory underlying the gravity solution, as well as the field
theory.)  As reviewed in section \ref{sec24}, this is best done in the
type IIA language, where it is clear that the Page charges are
quantized.  In particular, note that the parameter $q$ appearing in
(\ref{ansatz3}) does not intrinsically need to be quantized, as it is
a Maxwell charge, not a Page charge.

We wish to reduce from M-theory to IIA in such a way that the
infrared geometry is $AdS_4 \times CP^3$, so in particular we need
to have the $Z_k$ quotient act only on the M-theory circle
fibration. In terms of the original LWY coordinates, this means that
the M-theory circle should be the circle fibration orthogonal to the
1-form $(d\varphi_1 + A_1)$.  Explicitly,
\beq
\sigma = \left(
\frac{d\varphi_2 + A_2}{k} -\frac{U_2}{U_1+U_2} (d\varphi_1 + A_1)
\right)
\eeq
and
\beq
J=d\sigma = \frac{dA_2}{k}- \frac{U_2}{U_1+U_2} dA_1 -
d\left(\frac{U_2}{U_1+U_2}\right) \wedge (d\varphi_1 + A_1)
\eeq
reduces to the Kahler form on $CP^3$ in the deep infrared (with a
normalization which matches the conventions of
\cite{Aharony:2008ug}.) It is also useful to define a coordinate
$\rho$ by
\beq
w_1= \rho^2 \cos^2 \xi, \qquad w_2 = \rho^2 \sin^2 \xi\ .
\eeq

Now we can follow the recipe described in Section
\ref{sec24} to compute the D4-brane Page charge\footnote{
This procedure requires us to integrate over a cycle in the homology
class of the finite $CP^2$ in $CP^3$ at the bottom of the throat.
One parametrization of this cycle in our coordinates is
$\theta_1=\phi_1=0$ and $\rho=$ fixed.\label{footnotecp2}}.
Comparing with (\ref{c3}), we make the identification
\beq
q =-{m \over 4\pi g_s l_s^3}
\eeq
and write
\be C_3 =dt\wedge dx^1 \wedge dx^2 H^{-1}+ {m \over 8 R_1^2 (kR_2)^2}
 P^1 \wedge \frac{2}{U_2} \left(d\varphi_2'+A_2'\right)
+ \alpha J \wedge d \varphi_{11} \ , \label{c3tntn}
\ee
where we have added the potential term proportional to $\alpha$.
Because the Page charge is conserved and localized, we can compute it
at any convenient value of $\rho$, and choosing $\rho=0$, the
computation simply reduces to the ABJ case reviewed in Section
\ref{sec24}, and the result carries over.  The D4 Page charge is then
quantized as
\be (2 \pi)^2 \alpha = -(2 \pi l_s)^3 g_s \left(l - {k \over 2} \right). \label{alphatntn}\ee

The D4-brane Maxwell charge, in contrast with the Page charge, is
only truly well-defined at infinity. At generic values of $\rho$,
however, one can still define a radially-dependent effective D4
Maxwell charge by integrating $\tilde{F}_4$ (equivalently, $G_4$ in
M-theory) over a $CP^2$ cycle\footnote{To be explicit, one can
choose the same parametrization in footnote \ref{footnotecp2}.  For
small $\rho$ the 8-d geometry before including warping is $R^8/Z_k$
and the cycle is a  $CP^2$ with Fubini-Study metric; for large
$\rho$ the geometry deviates from the flat orbifold and the induced
metric on the 4-cycle will no longer be Fubini-Study.}. Given the
form of (\ref{c3tntn}), we have
\beq
Q_4^{Maxwell}(\rho)=\frac{1}{g_s (2\pi l_s)^3}  \int_{CP^2(\rho)}
(-\tilde{F}_4) = k b(\rho) + l-{k \over 2}.
\eeq
where
\be b(\rho) = {1 \over (2 \pi l_s)^2} \int_{CP^1} B_2.
\ee
It can be shown that $b(\rho)$ takes the form
\be b(\rho) = b_\infty f(\rho) - {(l-{k \over 2}) \over k} (1 - f(\rho))
\ee
where
\be f(\rho) =  \int d \xi \,
  \frac{2 \rho ^4 \cos \xi \sin ^3\xi R_1^2}{\left(\rho ^2 \cos ^2 \xi
    +R_1^2\right){}^2 \left(\rho ^2 \sin ^2\xi+(kR_2)^2\right)}\label{frho}
\ee
is a function with smoothly interpolates from $f(\rho)=0$ for $\rho=0$
and $f(\rho)=1$ for $\rho = \infty$. This feature is independent of
the values of $R_1$ and $R_2$ as long as they are finite.

We also see that
\be q = Q_4^{Maxwell}(\rho=\infty)\ . \ee
Interestingly, the asymptotic value of the D4 Maxwell charge is
independent of whether one integrates over a $CP^2$ or $S^2 \times
S^2$ cycle.

\begin{figure}
\centerline{\includegraphics[width=3in]{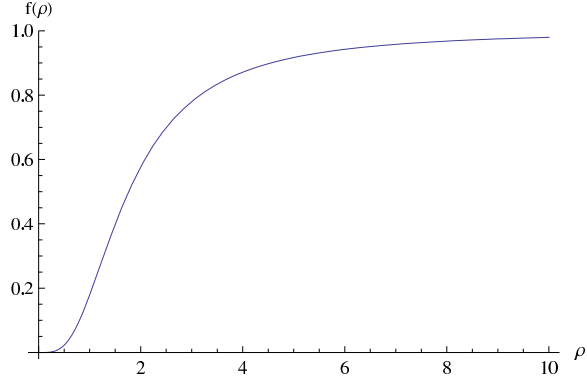}}
\caption{Numerical plot of $f(\rho)$ defined as an integral expression in (\ref{frho}). The fact that the function interplates from 0 at $\rho=0$ to $1$ at large $\rho$ is independent of the values of $R_1$ and $R_2$.
\label{fige}}
\end{figure}

\subsection{2-Brane Charge, Supersymmetry, and Singularities}

In the presence of the self-dual four-form flux, the supergravity
solution contains an induced 2-brane charge whose value at infinity
is:
\beq
\frac{1}{2(2\pi l_p)^6} \int_{{\cal M}_8} G_4 \wedge G_4
= \frac{q^2}{2k}.
\eeq
Thus the total 2-brane charge is given by
\beq
Q_2^{Maxwell}(\infty) = Q_2^{Maxwell}(0)+\frac{\left(k b_\infty + l-{k \over
2}\right)^2}{2k}.
\eeq
At generic values of $\rho$, we can also define an effective D2
charge obtained by integrating $* \tilde{F_4}$ over a surface at
fixed $\rho$.  This gives
\be Q_2^{Maxwell}(\rho) = \left(N + { k \over 8}\right) +  \left(l - {k \over 2}\right) b(\rho) + {k \over 2} b(\rho)^2 \ . \ee
The fact that the 2-brane charge varies as a function of $\rho$
should have an interpretation as a variation in the rank of the
gauge group with RG scale.

Because we know $b(\rho)$ for all $\rho$, we also know how
$Q_2^{Maxwell}(\rho)$ behaves near $\rho=0$. It is
\be Q_2^{Maxwell}(\rho=0)=  N - {l(l-k) \over 2k} \label{q2n4core}\ee
and is precisely the radius of $AdS_4$ geometry computed for the ABJM
model \cite{Bergman:2009zh} (excluding the contribution from higher
curvature corrections.) This is not at all surprising in light of the
fact that this ${\cal N}=4$ construction approaches the warped $R^8
/Z_k$ near the core which is precisely the ABJM geometry. It should
also come as no surprise that $Q_2^{Maxwell}(0)$, being related to the
AdS radius in the IR, is a gauge invariant combination of Page
charges.\footnote{What we call $Q_2^{Maxwell}(0)$ is equivalent to
what is more commonly referred to as the brane charge
\cite{Marolf:2000cb}.} What is intriguing about this result is the
fact that the positivity of $Q_2^{Maxwell}(0)$ imposes the same
condition as the condition for supersymmetry inferred from the
$s$-rule (\ref{srule}).

On the gravity side, the interpretation of the positivity of
$Q_2^{Maxwell}(0)$ is simple.  Given a set of UV brane charges,
naively one can construct a supersymmetric supergravity solution, but
for some values of the charge data $N$, $l$, and $k$,
$Q_2^{Maxwell}(\rho)$ must change sign from being positive to negative
at some $\rho$.  When $Q_2$ changes sign, the warp factor vanishes and
the geometry has a naked singularity of repulson type
\cite{Johnson:1999qt}.

Alternatively, consider the ``threshold'' case,
\be N = {l (l-k) \over 2k} \ee
for which $Q_2^{Maxwell}(0)=0$.  The gravity solution for this case
can be obtained by extrapolating from the supersymmetric solutions
with positive $Q_2^{Maxwell}(0)=0$, although the solution might have
large curvature corrections in the deep core region.  Now, if we add
anti-D2-branes to this threshold system, we will obtain a
non-supersymmetric solution with negative $Q_2^{Maxwell}$.  (The
threshold system has no ordinary D2-charge at its core so the
anti-branes have nothing to annihilate with; thus one might expect
this system to be metastable. We are ultimately interested in gravity
description of the true vacuum of this system with the same quantum
number as the aforementioned configuration with anti D2-brane.)  In
this regard, the nature of the threshold case is very similar to the
construction of Maldacena and Nastase \cite{Maldacena:2001pb}.

The generic feature of the supersymmetry breaking solution and its
implication for the dual gauge field theory for this construction is
expected to be similar to the ${\cal N}=3$ case. It would be
interesting to further explore the IR dynamics of the threshold
solution.

\begin{figure}
\centerline{\includegraphics[width=3in]{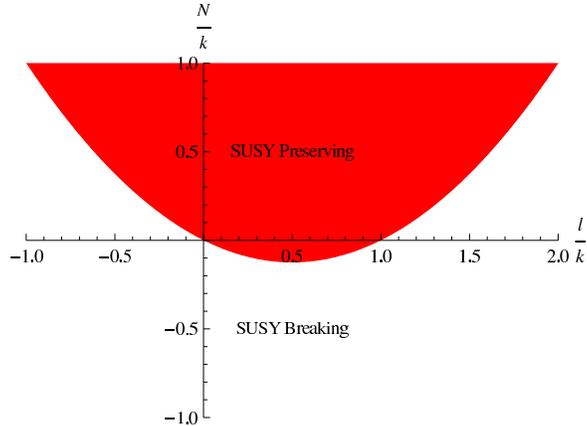}}
\caption{The red parabola indicates the range of $N$ and $l$ for which the system is expected to preserve supersymmetry.
\label{figf}}
\end{figure}

The condition on $N$, $l$, and $k$ for supersymmetry is a discrete
relation.  However, for the purpose of illustration, one can imagine
parameterizing
\be N = x k, \qquad l = y k\ee
in which case the condition for supersymmetry becomes
\be x > {y(y-1) \over 2} \ee
which is a parabola. The distinct physical configurations in an
interval of $x$ and $y$ grows when $k$ is taken to be a large
integer. By parameterizing the charges in terms of $x$ and $y$ for
large $k$, we arrive at an effectively continuous description of the
space of parameters of these field theories, with a threshold for
supersymmetry characterized by a region bounded by a smooth
parabola. The condition required for supersymmetry can be
represented visually in a simple phase diagram illustrated in figure
\ref{figf}.

It would be very interesting to better understand the IR situation
with broken supersymmetry.  On the field theory side, it appears to
be a dynamical supersymmetry breaking.  At least in the ${\cal N}=3$
case, the UV field theory has a definition in terms of a
supersymmetric Lagrangian.  If this supersymmetry breaking is
captured by a gravity dual, one might expect the existence of a
smooth solution in the infrared carrying anti-brane charge,
resolving the naked singularity which we identified.  A closely
related problem is to understand the supergravity solution with
explicit anti-branes, along the lines of
\cite{Brax:2000cf,DeWolfe:2008zy,Bena:2009xk}.  Either task is
extremely challenging technically, and involves solving a high-order
system of non-linear partial differential equations in two
variables.

\subsection{Discussion}

Although it was not necessary for the preceding analysis, to fully
specify the supergravity solution, we need to compute the warp
factor $H$. In the presence of background four-form flux, the
equation for $H$ acquires an inhomogeneous term. To handle this term
it is most efficient to use the Green's function derived in Appendix
\ref{greensfn}.  The answer in integral form is
\beq
H= N G(\vec{w}_1,0;\vec{w}_2,0) + \int d^3\vec{w}_1' d^3\vec{w}_2'
G(\vec{w}_1,\vec{w}_1';\vec{w}_2,\vec{w}_2')
v(\vec{w}_1',\vec{w}_2')
\label{warpfactor}
\eeq
up to an additive constant which we discard in the near-horizon
limit, and $v$ is the local induced 2-brane charge density,
\beq
v = q^2 {\pi^2 l_p^6 \over R_1^4 (kR_2)^4} {1 \over (U_1 w_1 U_2
w_2)^4}.
\eeq

The Green's function used to determine this warp factor has an
extremely complicated form (and we have not been successful in
attempts to simplify it.) Fortunately, in the preceding analysis
many features of the supergravity solution and the dual field theory
were understood without using the detailed form of $H$.

The supergravity solution we have presented here, including the warp
factor, is dual to a poorly understood field theory with 8
supercharges.  The background is somewhat difficult to work with,
because it is not conical -- the warp factor depends on two
variables instead of a single radial variable.

One can take the limit $R_1 \rightarrow \infty$ if one keeps
\be \rho_1 = \sqrt{2 R_1 r_1} = \sqrt{ w_1}  \ee
fixed, as was done in \cite{Hashimoto:2008iv} for the ${\cal N}=3$.
This geometry is presumably dual to some dynamical system in $2+1d$
although it is difficult to provide any alternative description for
it.

Our supergravity solution is related by T-duality to a configuration
of D5-NS5-D3-branes in type IIB, smeared along one direction.  This
type of triple brane intersection was studied by Lunin
\cite{Lunin:2007mj}.  There the (very nonlinear) supergravity
equations corresponding to this system were found and perturbative
solutions obtained. In our calculation we have exhibited an exact
solution to Lunin's equations, for one particular smearing.  It
would be interesting if our result gave some hints for the
construction of more general solutions preserving eight
supercharges.

\section{Holographic RG flow from $spin(7)$ holonomy manifold $A_8$}
\label{sec4}

In the previous section, we provided an embedding of the ABJM theory
in an RG flow with a UV fixed point whose gravity description was
simpler than the RG flow with a ${\cal N}=3$ Yang-Mills-Chern-Simons
UV fixed point.  We could find an explicit supergravity solution in
the range of parameters which preserves ${\cal N}=4$ supersymmetry,
and extrapolate up to the threshold of breaking of supersymmetry. In
order to find the solution beyond this threshold, however, even the
simplified ${\cal N}=4$ system appears to be too complicated.

It turns out that there are other known supergravity solutions which
can flow to ABJM.  Some of these solutions have large global
symmetry groups which can simplify the gravity analysis (at the cost
of reducing the supersymmetry, which makes the field theory even
more obscure.)  One such construction is the asymptotically locally
conical (ALC) geometry of $spin(7)$ holonomy originally constructed
by Cvetic, Gibbons, Lu, and Pope in \cite{Cvetic:2001pga}.
Compactification of M-theory on $spin(7)$ manifolds gives rise to a
gravity dual of a $2+1d$ dynamical system with ${\cal N}=1$
supersymmetry. These authors constructed explicit metrics for two
broad classes of $spin(7)$ holonomy manifolds which they called
$A_8$ and $B_8$. In this section we will focus on $A_8$ which is
more relevant in connection with the ABJM theory. We will provide
some additional discussion on $B_8$ in the next section.

The starting point in the construction of this $spin(7)$ manifold is
a construction of an explicit ansatz following the template of the
earlier work on $G_2$ holonomy manifolds
\cite{Bryant:1989mv,Gibbons:1989er}. Consider an ansatz for ${\cal
M}_8$ of the form
\be ds_{A_8}^2 = h(r)^2 dr^2  + a(r)^2  (D \mu^i)^2  + b(r)^2 \sigma^2 + c(r)^2d \Omega_4 \label{a8ansatz}\ee
where $\sigma^2$ and $(D \mu^i)^2$ are line elements of the $S^3$
fiber on $S^4$ base where $S^3$ itself is viewed as a $S^1$ fiber
over an $S^2$ base \cite{Cvetic:2001pga}. Through explicit
substitution to the equations of motion, one confirms easily that
the following is a solution.
\beq h(r)^2 & = & {(r+\ell)^2 \over (r+3 \ell)(r - \ell)} \cr
a(r)^2 & = & {1 \over 4}(r+3 \ell) (r - \ell) \cr
b(r)^2 & = & {\ell ^2 (r + 3 \ell)(r - \ell) \over (r+\ell)^2} \cr
c(r)^2 & = & {1 \over 2} (r^2 -\ell^2)
\eeq
The parameter $\ell$ is taken to be positive.  Topologically, this
space is $R^8$.  Although the metric appears to have a coordinate
singularity at $r=\ell$, the space is actually locally $R^8$ and
therefore regular.  The fact that $b(r)$ approaches a constant implies
that the geometry is asymptotically that of a product of a cone and an
$S^1$, i.e. the space is asymptotically locally conical
(ALC). Orbifolding the $S^1$ by $Z_k$ gives rise to an ALC whose core
is the $R^8/Z_k$.

Just as in the previous section, $\sigma$ can be written in the form
\be \sigma = d \varphi + {\cal A} \ee
where the $\varphi$ has period $4 \pi/k$. Using this $S^1$ to reduce
from M-theory to IIA, we find that
\be \ell = { k \over 2} g_s l_s \ . \ee

It is not too difficult to consider adding $N$ M2-branes by
considering an ansatz of the form (\ref{ansatz1}). As usual, one can
consider taking the near horizon limit by scaling $l_s \rightarrow 0$
keeping
\be U = {r \over l_s^2}, \qquad g_{YM2}^2 = {g_s \over  l_s} \ee
fixed, which gives rise to a geometry dual to some theory decoupled
from gravity.

One disadvantage of the $A_8$ construction compared to the ${\cal
N}=3$ construction is the fact that the UV fixed point of the
decoupled field theory is difficult to identify. The facts that the
amount of supersymmetry is ${\cal N}=1$ and that the IR physics is
in the same universality class as the ABJM theory suggest that this
system arises from configuration labeled 4(i) in table
\ref{tablea} with $\theta_{1,2,3}=2 \pi/3 + \tan^{-1}(k)/3$. Some
discussion about the field theory duals of these construction also
appears in
\cite{Loewy:2002hu} although it is harder to confirm these conjectures
in detail due to the small number of supersymmetries.  We will not
dwell further on this point, as it does not affect our gravity-side
analysis, but it would certainly be interesting to understand the
field theory better.

One helpful feature of this system is that the self dual 4-form on
$A_8$ is known explicitly.\footnote{In \cite{Cvetic:2001pga}, the
supersymmetry preserving 4-form on $A_8$ is referred to as being
anti-self-dual.  This convention requires that $b(r)$ is taken to be
negative, as explained briefly in footnote 4 of
\cite{Cvetic:2001pga}. We will instead adopt the convention where
$b(r)$ is positive, and for the supersymmetry preserving 4-form to be
self-dual, in order to match the conventions adopted in
\cite{Aharony:2009fc}. The fact that the supersymmetry preserving
4-form for the $B_8$ is anti-self-dual in the same set of conventions
(where $b(r)$ is positive), however, will turn out to be important in
Section \ref{sec5}.} It can be written in the form
\be C_3 =   m B_{(3)}+\alpha d \sigma \wedge d \varphi \ee
where
\be B_{(3)} = (r-\ell)^2 \left[ -{1 \over 8 (r + \ell)^2} \sigma \wedge X_{(2)} + {(r + 5 \ell) \over 8 (r+\ell)(r + 3 \ell)^2} \sigma \wedge Y_{(2)} - {1 \over 16 (r+3 \ell)^2} Y_{(3)} \right] \ee
as is given in \cite{Cvetic:2001pga}. We have also included a term
proportional to $\alpha$ which contributes only as an additive
constant to the NSNS B-field in the type IIA reduction. In fact, the
components of this 3-form which reduce to the NSNS 2-form are
\be B_2 = {2 \over k R}  m (r-\ell)^2 \left(-{1 \over 8 (r+\ell)^2} X_{(2)} + {(r+5\ell) \over 8 (r+\ell) (r+3\ell)^2} Y_{(2)} \right) + {2 \over k R} \alpha (X_{(2)} - Y_{(2)}) \ . \ee
In the type IIA picture, the surface of fixed $r$ is $CP^3$. Thus,
topology of the fixed $r$ slice of the geometry is the same in all
examples considered so far.  Then, $X_{(2)}$, $Y_{(2)}$, and
$Y_{(3)}$ are differential forms on $CP^3$ whose details can be
found in
\cite{Cvetic:2001pga}.

As was the case in the previous section, $m$ and $\alpha$ are
determined by the quantized D4 Page charge and the asymptotic value
of $b_\infty$. Specifically, they come out as
\be m =(4 \pi g_s l_s^3)  \left(-l + {k \over 2} - b_\infty k\right) , \qquad \alpha = - {(2 \pi l_s)^3 g_s \over 16 \pi^2} \left(l - {k \over 2 }\right) \ . \ee
Note, as in the previous case, that it is $\alpha$ which is
discretized, whereas $m$ is a continuous parameter, somewhat counter
to the naive expectation. With the values of $m$ and $\alpha$
determined in terms of the field theory data accordingly, the NSNS
2-form is such that\footnote{We warn the reader the symbol ``$b(r)$''
are used in two different contexts, one as the part of the line
element in CGLP ansatz (\ref{a8ansatz}), and other as the
dimensionless period integral of $B_2$ over $CP^1$
(\ref{bofr}). Hopefully the difference is clear from the context.}
\be b(r) = {1 \over (2 \pi l_s)^2} \int_{CP^1} B = b_\infty f(r) - {(l - {k \over 2}) \over k} (1 - f(r)) \label{bofr} \ee
for
\be f(r) = {r- \ell \over r+ \ell}\,  \ee
where once again, $f(r)$ is a function which smoothly interpolates from 0 to 1 as $r$ runs from $r=\ell$ to $r=\infty$.

Most of the conclusions concerning the condition for supersymmetry
follow from these observations. The Maxwell flux at fixed $r$ is
\be Q_2^{Maxwell}(r) = \left(N + { k \over 8}\right) +  \left(l - {k \over 2}\right) b(r) + {k \over 2} b(r)^2 \ee
which near the tip $r=\ell$ becomes
\be Q_2^{Maxwell}(r=\ell)=  N - {l(l-k) \over 2k} \label{q2a8core} \ . \ee
The interpretation of this formula is the same as before. If the
parameters $N$, $l$, and $k$ are chosen so that $Q_2^{Maxwell}(r)$
changes sign somewhere in the range $\ell < r < \infty$, then the
geometry has a naked singularity, precisely when supersymmetry is
expected to be broken. The fact (\ref{q2a8core}) is identical to
(\ref{q2n4core}) implies that the region of parameter space for SUSY
is also given by what is illustrated in figure
\ref{figf} for the model based on $A_8$.

The main reason for considering $A_8$ in addition to the ${\cal N}=4$
construction of the previous section is the fact that the $A_8$
background depends explicitly only on a single radial variable
$r$. This provides a simpler context to explore the supergravity
solution for the range of parameters where the supersymmetry is
expected to be broken where an ansatz more general than the BPS case
needs to be constructed. The form of the warped $A_8$ metric provides
a natural context to formulate such an ansatz.

Let us consider taking
\beq ds^2 &=& H^{-2/3} (-dt^2+dx_1^2 + dx_2^2) + H^{1/3} ds_8^2 \cr
F_4 & = & dt \wedge dx_1 \wedge dx_2 \wedge d \tilde H^{-1} + m G_{4}
\eeq
where
\beq ds_8^2 &=& h(r)^2 dr^2 + a(r)^2 (D \vec \mu)^2 + b(r)^2 \sigma^2 + c(r)^2 d \Omega_4^2 \\
C_3 &=& m \left( v_1(r) \sigma \wedge X_2 + v_2(r) \sigma \wedge Y_2 + v_3(r) Y_3 \right) + \alpha d \sigma \wedge d x_{11} \label{3formansatz}\eeq

This is the generic ansatz preserving the $SO(5)$ global symmetry of
the UV theory, so it is a reasonable guess that it contains the
non-supersymmetric solution as well, assuming that the gravity dual
exists and does not spontaneously break the global symmetry.  The BPS
ansatz assumed that $G_4$ is self-dual and that $H(r) = \tilde H(r)$
\cite{Cvetic:2001pga}.  To construct a non-supersymmetric solution we
should relax both of these requirements. This would then mean that
there are, in total, eight scalar functions $a(r)$, $b(r)$, $c(r)$,
$v_1(r)$, $v_2(r)$, $v_3(r)$, $H(r)$, and $\tilde H(r)$, satisfying
second order non-linear differential equations. (One of the functions
appearing in the ansatz $h(r)$ can be set to take on arbitrary form by
re-parametrization invariance of the $r$ coordinate.)

Solving this system of equations is still a challenging enterprise,
but at least in principle it can be done numerically if it is
supplemented with appropriate boundary conditions. A similar
analysis in the context of Klebanov-Strassler system was carried out
recently in \cite{Bena:2009xk}.  In the UV, the boundary conditions
are imposed by demanding that the fluxes are characterized by the
quantum numbers $N$, $l$, $k$, and $b_\infty$.  In the IR, we seek a
smooth solution, so the eight scalar functions should approach
constants.  This is still not a totally satisfactory formulation,
but it is a significant improvement over the ${\cal N}=4$
construction, provided of course that the ansatz being proposed is
general enough to contain the solution we are ultimately after. Some
of the key features to extract from the SUSY breaking solution are
the vacuum energy and dependence on coordinates in figure
\ref{figf}, which we expect to read off from the ADM mass along the
lines of
\cite{Maldacena:2001pb,DeWolfe:2008zy}.\footnote{An interesting class
of non-BPS deformation interpolating $A_8$ and the 8 dimensional
Taub-NUT solution was discovered in \cite{Bizon:2007zf}. The near
horizon limit of these geometries including the back reaction of D2
branes presumably gives rise to gravity dual of some non-BPS
deformation of the decoupled field theory discussed in this section.}
\footnote{A different class of non-supersymmetric solution, corresponding to finite temperature generalization of a similar construction, was considered in \cite{Giecold:2009wj}.}

\section{Holographic RG flow from $spin(7)$ holonomy manifold $B_8$}
\label{sec5}

In addition to the analysis of supersymmetry breaking for a theory
related to ABJM in the previous section, there are some additional
interesting observations one can make in a closely related
construction. In this section, we will discuss one broad class of
constructions based on the $spin(7)$ holonomy manifold called $B_8$ in
\cite{Cvetic:2001pga}, which gives rise to a gravitational dual of
${\cal N}=1$ Chern-Simons theory \cite{Gukov:2001hf}. There are two
main observations in this section.  The first is that for the values
of field theory parameters which lead to dynamical supersymmetry
breaking \cite{Witten:1999ds}, the gravity solution has a naked
singularity as in the previous section.  The other observation
concerns the fact that for the $B_8$ background, Cvetic
et.al. succeeded in constructing a solution to the equation of motion
corresponding to a deformation through fluxes which breaks all of the
supersymmetries. We will show that this solution covers some, but not
all, of the range of parameters where the supersymmetry is broken.

\subsection{Review of construction of $B_8$ manifolds}

Let us begin by reviewing the construction of the $B_8$ manifolds
which are eight dimensional manifolds with $spin(7)$ holonomy group.
The topology of $B_8$ is that of a spin bundle over $S^4$ (in contrast
with $A_8$ which was topologically trivial.)

The simplest example of
such a $spin(7)$ holonomy space with this
topology
is the asymptotically conical manifold of
\cite{Bryant:1989mv,Gibbons:1989er} which has the form
\be ds_8^2 = \left(1 - {\ell^{10/3} \over r^{10/3}}\right)^{-1} dr +{9 \over 100} r^2 \left(1 - {\ell^{10/3} \over r^{10/3}} \right) h_i^2 +{9 \over 20} r^2 d \Omega_4 \ , \label{ac} \ee
with
\be h_i \equiv \sigma_i - A^i_{(1)} \ , \ee
where $\sigma_i$ are left invariant one-forms on $SU(2)$,  and
$A^i_{(1)}$ are $SU(2)$ Yang-Mills instanton on $S^4$. For $\ell =
0$, this geometry reduces to a cone whose base is a squashed $S^7$
\cite{Duff:1983nu,Duff:1986hr}. The case with finite $\ell$
corresponds to deforming the tip of this cone so that there is a $S^4$
of finite radius at $r=\ell$.

There is also a slightly more complicated example, called the $B_8$
manifold.  Its metric
may be obtained from the metric of $A_8$ by taking the parameter
$\ell$ to be negative so that in terms of the positive quantity
$\tilde
\ell= |\ell|$, the $B_8$ metric has the form
\be ds_8^2 = {(r -\tilde \ell) ^2 \over (r - 3 \tilde \ell) (r +
\tilde \ell)} dr^2
+ {1 \over 4} (r - 3 \tilde \ell) (r + \tilde \ell) (D \mu^i)^2
+ {\tilde \ell^2 (r - 3 \tilde \ell) (r + \tilde
\ell) \over (r - \tilde \ell)^2} \sigma^2
+ {1 \over 2} (r^2 - \tilde \ell^2) d \Omega_4^2 \label{b8} \ . \ee
This extrapolation essentially amounts to analytically continuing the
radius of the $S^1$ of the $A_8$ to negative value. Often, such an
extrapolation gives rise to a singularity, but in this case the
solution is perfectly regular.  In the infrared, the geometry
approaches $R^4 \times S^4$, and as such is in the same universality
class as the deformed solution (\ref{ac}). Just as in the case of
$A_8$, we can quotient this geometry by the action of $Z_k$ subgroup
of this $S^1$.  The solution (\ref{b8}) can further be understood as a
point in one dimensional family of solutions, named $B_{8+}$ and
$B_{8-}$ in \cite{Cvetic:2001pga}, where one varies the ratio of the
radius of the $S^4$ at the tip to the asymptotic radius of $S^1$ at
large $r$.  When the radius of $S^4$ is pushed to zero, the geometry
is asymptotically locally conical, just like the $B_8$, which
interpolates to an undeformed cone whose base is a squashed $S^7$
\cite{Duff:1983nu,Duff:1986hr}.  (This was studied in the holographic
context in \cite{Ooguri:2008dk}.)  We call this geometry
$B_{8\infty}$. By making the radius of $S^4$ finite keeping the radius
of $S^1$ fixed, we are deforming the IR of the $B_{8\infty}$ just like
$\ell$ in (\ref{ac}) is deforming the tip of the squashed $S^7$ cone.

\begin{figure}
\setlength{\unitlength}{3947sp}%
\begingroup\makeatletter\ifx\SetFigFont\undefined%
\gdef\SetFigFont#1#2#3#4#5{%
  \reset@font\fontsize{#1}{#2pt}%
  \fontfamily{#3}\fontseries{#4}\fontshape{#5}%
  \selectfont}%
\fi\endgroup%
\centerline{\begin{picture}(3357,529)(2656,-2378)
\thinlines
{\color[rgb]{0,0,0}\put(2701,-1861){\line( 0,-1){300}}
}%
{\color[rgb]{0,0,0}\put(4201,-1861){\line( 0,-1){300}}
}%
{\color[rgb]{0,0,0}\put(2701,-2011){\vector( 1, 0){3300}}
}%
\put(2656,-2356){\makebox(0,0)[lb]{\smash{{\SetFigFont{12}{14.4}{\rmdefault}{\mddefault}{\updefault}{\color[rgb]{0,0,0}$B_{8\infty}$}%
}}}}
\put(3386,-2378){\makebox(0,0)[lb]{\smash{{\SetFigFont{12}{14.4}{\rmdefault}{\mddefault}{\updefault}{\color[rgb]{0,0,0}$B_{8+}$}%
}}}}
\put(4136,-2378){\makebox(0,0)[lb]{\smash{{\SetFigFont{12}{14.4}{\rmdefault}{\mddefault}{\updefault}{\color[rgb]{0,0,0}$B_8$}%
}}}}
\put(5056,-2378){\makebox(0,0)[lb]{\smash{{\SetFigFont{12}{14.4}{\rmdefault}{\mddefault}{\updefault}{\color[rgb]{0,0,0}$B_{8-}$}%
}}}}
\put(6100,-2080){\makebox(0,0)[lb]{\smash{{\SetFigFont{12}{14.4}{\rmdefault}{\mddefault}{\updefault}{\color[rgb]{0,0,0}$\lambda$}%
}}}}
\end{picture}%
}
\caption{Schematic illustration of one parameter ($\lambda$) family of
deformation of $B_8$ geometry. All of these geometries have identical
asymptotic UV geometry. At $\lambda =\lambda_\infty$, the geometry
asymptotes to a form interpolating between $B_8$ in the far UV, and
$AdS_4 \times S^7_{squashed}$ in the IR.\label{figg}}
\end{figure}
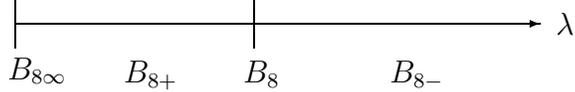

The one parameter family of $B_{8\pm \infty}$ solutions can be
summarized in a diagram illustrated in figure \ref{figg}. A similar
diagram appears in figure 3 of \cite{Cvetic:2001zx} where the
horizontal axis is interpreted as varying the radius of $S^1$
keeping the size of $S^4$ fixed. The explicit metrics for $B_{8+}$,
$B_{8-}$, and $B_{8\infty}$ are somewhat cumbersome to write
explicitly. We will summarize some of these details in the appendix.

As in the $A_8$ geometry, the $B_8$ manifold supports normalizable
4-forms.  In the case of $B_8$, with the convention that $a(r)$,
$b(r)$, $c(r)$, and $h(r)$ are taken to be positive, it is the
anti-self-dual 4-form which leaves the supersymmetry unbroken
\cite{Cvetic:2001pga}. It is given by \be G_4 = d C_3 \ee
where
\be C_3 = m \left( v_1(r) \sigma \wedge X_2 + v_2(r) \sigma \wedge Y_2 + v_3(r) Y_3 \right) + \alpha d \sigma \wedge d \varphi \ee
and
\beq v_1(r) &=& - {r^5 + 5 \tilde \ell r^4 + 10 \tilde \ell^2 r^3  + 10 \tilde \ell^3 r^2 - 155\tilde \ell^4 r + 97\tilde \ell^5 \over 8 (r+\tilde \ell)^3 (r-\tilde \ell)^2} \cr
v_2(r) & = & - {r^4+6 \tilde \ell r^3 + 12 \tilde \ell^2 r^2 - 22 \tilde \ell^3 r + 35\tilde \ell^4  \over 8 (r-\tilde \ell)(r+\tilde \ell)^3} \cr
v_3(r) & = & - {r^3 + 11 \tilde \ell r^2 + 67 \tilde \ell^2 r - 7 \tilde \ell^3\over 16 (r+\tilde \ell)^3} \ . \label{b8asd}
\eeq
Just as in the case for $A_8$, the values of $m$ and $\alpha$ are
determined by fixing $b_\infty$ and $l$.  It takes the standard form
\beq  m &=&  - (4 \pi g_s l_s^3 )\left(l - {k \over 2}+b_\infty k \right) \\
(2 \pi)^2 \alpha &=&  -(2 \pi l_s)^3 g_s \left(l - {k \over 2} \right)\ .
\eeq

These conditions are obtained by imposing a
quantization condition on the D4 {\it Page} charge (\ref{pageflux})
through the $S^4$ cycle in the type IIA description of the $B_8$
geometry.\footnote{Because of the identity $(X_2 - Y_2) \wedge (X_2
- Y_2) = 6 \Omega_4 - d Y_3$, it follows that $\int_{CP^2} (X_2 -
Y_2) \wedge (X_2 - Y_2) = \int_{S^4} (X_2 - Y_2) \wedge (X_2 - Y_2)
= 16 \pi^2$ and so the quantization of Page flux through $CP^2$ and
$S^4$ gives rise to the same discretization condition on $\alpha$.}
This is physically distinct from imposing a quantization condition
on the M-theory 4-form flux through the $S^4$ as is done, e.g., in
\cite{Gukov:2001hf,Martelli:2009ga}. The latter approach amounts to
imposing a quantization condition on the {\it Maxwell} charge, and
gives rise to some subtle difference between the two approaches in
the indentification of parameters of the supergravity background and
its gauge theory dual.

For BPS solutions, the warp factor due to the presence of D2-brane
charges can be determined as a solution to the Laplace equation as
was described in \cite{Cvetic:2001pga}.  The simplest case to
consider is an ansatz where the warp factor is uniform along
surfaces of fixed $r$. Such an ansatz is appropriate when large
number of D2 charges are distributed uniformly on the $S^4$ near the
core. It is not too difficult, though somewhat cumbersome, to relax
this ansatz and solve for the harmonic functions with less
symmetries. The condition on parameters such as fluxes and charges
necessary for supersymmetry will be reviewed in the remainder of
this section.

There is one more interesting feature about the $B_8$ which is
different from the LWY, $TN\times TN$, and the $A_8$ which we
considered earlier: On $B_8$, there also exists a normalizable {\it
self-dual} 4-form which breaks all supersymmetry whose analytic form
is known \cite{Cvetic:2001pga}. It is given by taking
\beq v_1(r) & = & -{5 r^4 - 20 \tilde \ell r^3 + 38 \tilde \ell^2 r^2 - 36 \tilde \ell^3 r + 29\tilde \ell^4 \over 8 (r-\tilde \ell)^2 (r+\tilde \ell)^2}\cr
v_2(r) & = & - {5 r^3 - 15 \tilde \ell r^2 + 19 \tilde \ell^2 r + 7\tilde \ell^3 \over 8 (r-\tilde \ell) (r+\tilde \ell)^2} \cr
v_3(r) & = & - { 5 r^2 + 2 \tilde \ell r + 13 \tilde \ell^2\over 16 (r+\tilde \ell)^2} \label{b8sd}
\eeq
and the relation between $m$, $\alpha$, $b_\infty$, and $l$ take
similar forms to what we found in the anti-self-dual case.  The
physical implication of the existence of this self-dual 4-form will be
discussed below.

\subsection{$B_8$ manifold and the ${\cal N}=1$ Chern-Simons Theory}

The physical interpretation of the deep infrared dynamics of
M-theory compactified on $B_8$ is ${\cal N}=1$ pure Chern-Simons
theory with gauge group $SU(N)_k$ \cite{Gukov:2001hf}. This arises
if one considers taking the $Z_N$ orbifold along the $S^1$ and
reducing to IIA. Then, the RR 1-form arising from the fibration
along $S^1$ can be interpreted as $N_{GS}$ D6-branes wrapping the
$S^4$ cycle.

Then, on the world volume of the D6-branes, there will be an effective
Chern-Simons coupling due to the Wess-Zumino term
\beq S &=& \int_{R^{1,2} \times  S^4} {1 \over 2} (F+B) \wedge (F+B) \wedge A_3 +  {1 \over 6} (F+B) \wedge (F+B) \wedge (F+B) \wedge A_1 \cr
& = &\int_{R^{1,2}} A \wedge F \int_{S^4} (-d A_3 - H_3 \wedge A_1 - B_2 \wedge F_2) \cr
& = & k_{GS} \int_{R^{1,2}} A \wedge F
\eeq
where $k_{GS}$ is the discrete data corresponding to the Page flux
through $S^4$.

For the purpose of embedding ${\cal N}=1$ Chern-Simons theory as the
low energy effective physics, one can just as well work with the
deformed asymptotically cone background (\ref{ac}) where the
anti-self-dual 4-form is just as easy to find.

The data characterizing $B_8$ share some features with $TN \times
TN$ and $A_8$ but also have some crucial differences.  Just as in the
earlier discussions, The parameter $m$, however, which varied
continuously in the other cases, must be quantized in the $B_8$
background, because $B_8$ has a topologically nontrivial finite-sized
four-cycle, and the integral of $G_4$ on this cycle must be quantized:
\be \frac{1}{(2 \pi l_p)^3}\int_{\mathcal{M}_4} \left( G_4+
\frac{1}{16\pi} \tr R\wedge R \right)= q \ee with $q$ an integer.
Note that the quantization condition for $G_4$ on an eight-manifold
receives a correction associated with a multiple of the first
Pontryagin class, as described in \cite{Witten:1996md}.  It was shown
in \cite{Gukov:2001hf} that this contribution shifts the quantization
law for $G_4$ in $B_8$ by a half unit.  Specifically, for the
background given by (\ref{b8})--(\ref{b8asd}), the quantization
condition reads
\be \int_{S_4} G_4 = m u_1 c^4 \Omega_4 = 5 \pi^2 m = (2 \pi l_s)^3 g_s \left(q-{k \over 2} \right)\ee
or equivalently
\be -{5 \over 2} \left(l - {k \over 2} + b_\infty k\right) = q -{k \over 2}\label{binfq}\ee
where $u_1$, in the notation of (\ref{b18}), is a component of $G_4 =
d C_3$ along the $S_4$.  Equation (\ref{binfq}) implies, perhaps
somewhat surprisingly, that $b_\infty$ is constrained in this
geometry. This parameter can, however, be tuned by deforming the
(\ref{b8}) geometry as we describe in appendix \ref{appendixb}.

Adding D2-brane sources gives rise to additional warping. 
The sources for this warping include explicit D2's as well as induced D2-brane
charge on the world volume of D4 and D6 branes as a result of the
NS-NS B-field present in the background. Computing the Maxwell charge
as we did in earlier sections gives
\be Q_2^{Maxwell}(r = \infty) = N + {k \over 8} +\left( l - {k \over 2}\right)b_\infty + {k \over 2} b_\infty^2 \ . 
\ee
What is more important is the Maxwell charge at the tip $r=3 \tilde
\ell$ for which we find
\be Q_2^{Maxwell}(r=3 \tilde \ell) = 
N - {l (l - k) \over 2k } + {\left(q-{k \over 2}\right)^2 \over 2 k} \label{b8asdrf} \ee
where the last term comes from term proportional to $m^2$.  The
discussion of \cite{Cvetic:2001pga} focused primarily on the case
where $Q_2^{Maxwell}(r = 3 \tilde \ell)=0$ as the corresponding
gravity solution will have a singularity.  But this singularity can be
attributed to the presence of brane sources i.e. D6 wrapping the
$CP^2$ at the tip, so we will consider it physically allowed.

Let us now focus on the case where $q=[k/2]$ is the integer part of
$k/2$, and $N=0$ so that the IR dynamics do not contain any additional
dynamical degrees of freedom besides thpse of the ${\cal N}=1$
Chern-Simons theory. Because the supersymmetry preserving flux on
$B_8$ is anti-self-dual, it preserves the same supersymmetry as that
of anti-D2-branes. Adding a D2-brane will break supersymmetry. This
means that the condition for the absence of a repulson singularity is
\be Q_2^{Maxwell}(r = 3 \tilde \ell) = - {l(l - k) \over 2k } < 0
\ee
with the inequality pointing in the direction as indicated. We have
also dropped the possible contribution ${1/ 8 k}$, which only arises
if $k$ an odd integer, becuase it is subleading in $1/k$.

From this, we infer that
\be \left(l - {k \over 2}\right)   > {k \over 2} \ . 
\ee
Recall that $k$ in this context refers to the number of D6-branes,
which Gukov and Sparks denote $N_{GS}$. The combination $(l - k /2)$,
on the other hand, is the Page flux which in the language of Gukov and
Sparks is $k_{GS}$.\footnote{It is worth noting that the half-integer
quantization of the D4 Page charge is consistent with the half-integer
quantization of $k_{GS}$ due to the parity anomaly
\cite{AlvarezGaume:1983ig,Niemi:1983rq,Redlich:1983dv}.} So, the
repulson-free condition reads
\be k_{GS} > {N_{GS} \over 2} \ee
which is precisely the condition for the supersymmetry to be unbroken
according to the index computation of Witten \cite{Witten:1999ds}. We
therefore conclude that this general pattern of identifying the
threshold of supersymmetry from the appearance of a naked singularity
applies to the construction of Gukov and Sparks as well.

\subsection{$B_8$ with supersymmetry breaking 4-form flux}

Let us now consider the generalization of the previous section where
we allow $N$ and $b_\infty$ to take on general values.  Then, the
supergravity solution has a structure roughly resembling that of a
cascading gauge field theory which we saw in the earlier sections.

The repulson-free condition is the same as what we found in
(\ref{b8asdrf}), and so in terms of
\be x = {N \over k}, \qquad y = {l \over k} \ee
it reads
\be x < {y(y-1) \over 2}   - {25 \over 8} \left(y - {1 \over 2} + b_\infty\right)^2 \ee
which is a parabola pointing down.

On first pass, this is the end of the story, where, like in the
construction of previous sections, we must adopt a general non-BPS
ansatz to explore the solutions outside the range of parameters
covered by this parabola.

\begin{figure}
\centerline{\includegraphics[width=3in]{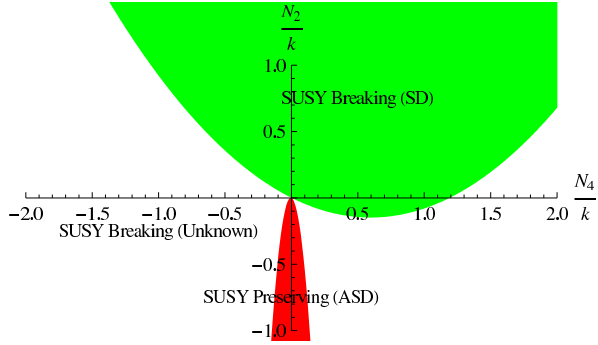}}
\caption{The red parabola indicates the region parameterized by $N$
and $l$ for the $B_8$ geometry in the presence of anti-self-dual
4-form field strength in the background where the supersymmetry is
expected to be unbroken. In the region outside the red parabola, the
supersymmetry is expected to be broken. The green parabola indicates
the region where although supersymmetry is expected to be broken,
there exists a dual gravity description in terms of $B_8$ geometry
with non-vanishing self-dual 4-form field strength.}  \label{figh}
\end{figure}

In the case of $B_8$ manifolds, however, there are known supergravity
solutions with normalizable self-dual 4-form flux (\ref{b8sd}) which
are not supersymmetric. It is natural to ask what role these solutions
play in capturing the range of parameters outside the supersymmetry
parabola.  The self-dual 4-form acts as bulk sources for positive
D2-brane charge. One therefore imagines that adding an anti-D2 will
have a more dramatic effect on this background through brane
annihilation, whereas adding a D2 should have a milder effect in light
of the fact that the supersymmetry is already broken by the presence
of the fluxes.\footnote{One can regard the resulting geometry as a
type of skew-whiffed geometry
\cite{Duff:1983nu,Duff:1986hr,Duff:1995wk}. Application of
skew-whiffing in the context of AdS/CFT correspondence was studied in
\cite{Berkooz:1998qp,Berkooz:1999ji}.} The Maxwell charge at the tip
for the self-dual 4-form flux can be computed and we find
\be Q_2^{Maxwell}(r=3 \tilde \ell) = N - {l (l - k) \over 2k } +
{4 \over 25 } {\left(l - {k \over 2} + b_\infty k\right)^2 \over 2k}
\label{b8sdrf} \ . \ee
The gravity solution will therefore contain repulson singularity unless
\be Q_2^{Maxwell}(r=3 \tilde \ell) > 0 \ . \ee
This will define another parabola. It is instructive to illustrate the
parabolic repulson-free region for the anti-self-dual and self-dual
4-forms on the same axis. For the sake of illustration, let us fix
$b_\infty = 1/2$ (which will constrain the value of $q$). The
resulting phase diagram is illustrated in figure \ref{figh}. The red
parabola indicates the repulson-free region for the background with
anti-self-dual 4-form and is expected to correspond to the region
where supersymmetry is unbroken. The green parabola indicates the
repulson-free region of the background with self-dual
4-form.\footnote{The edges of these parabolas are also special in that
the 2-brane source $Q_2^{Maxwell}(r=3 \tilde \ell)$ at the core is
zero, making the dual gravity completely regular.} Outside the two
parabolas, the solutions are not known, and we expect a more general
ansatz to be required in order to find them.

A curious point to note is that the two parabolas touch at one
point. This appears to be a generic feature which follows from the
fact that the charge at the tip takes the form
\be Q_2^{Maxwell}(r = 3 \tilde \ell) = N -{l(l - k) \over 2k} +  {C(\lambda) \over 2k} \left(l - {k \over 2} + b_\infty k\right)^2  \label{repulsonfree}\ee
where $C(\lambda)$ is some constant and $\lambda$ is the variable
parameterizing the family of $B_{8+}$, $B_{8-}$, $B_{8}$, and
$B_{8\infty}$ solutions illustrated in figure \ref{figg}.  The fact
that both self-dual and anti-self-dual 4-form give rise to brane
charge of this generic form appears to hold also for $B_{8+}$ and
$B_{8-}$.\footnote{We will elaborate further on this point in
Appendix \ref{appendixb}.} So for the entire class of
$B_{8\pm\infty}$ backgrounds, we find a repulson-free region in the
phase space parameterized by $x=N/k$ and $y=l/k$ which describes two
parabolas touching at one point.

This means that for generic deformation away from the supersymmetric
region indicated by the red parabola, one finds that one must apply
the generalized ansatz discussed at the end of section 4. But, at one
point along the supersymmetry breaking threshold, there is a
``bridge'' to a domain in phase-space where gravity solution can be
prescribed using a simpler ansatz, and its form is known. It might be
interesting to study this region of parameter space more closely.

\section{Conclusions}
\label{sec6}

In this article, we explored variety of cascading field theories in
2+1 dimensions from the point of view of the gravity dual. The
models we considered are UV embeddings of ABJM theory and their
cousins. We analyzed the quantization of charges of the gravity
solution, and identified their interpretation in terms of discrete
data, i.e.\ ranks and levels on the field theory side. The condition
that some amount of supersymmetry is left unbroken, which can be
understood as the arising from generalized $s$-rule in the brane
construction, manifests itself as the condition for the absence of a
certain type of singularity in the dual gravity description. For
LWY, $TN\times TN$, and $A_8$, this condition takes on a simple
gauge invariant form\footnote{This condition is modified slightly
for the $B_8$ models.}
\be N- {l(l-k) \over 2k} > 0 \ , \label{eq6.1}\ee
which when satisfied, flows to ABJM theory in the IR. These conditions
were represented by the ``red parabola'' in the $(N,l)$ space in
earlier sections. This is the basic result around which we build
further observations.

One obvious yet important question concerns the nature of low energy
effective dynamics of this system when the charges do not satisfy the
condition for preserving supersymmetry. While the naive extrapolation
of the dual gravity solution is singular, one expects this singularity
to be resolved by some mechanism.  This situation is strongly
reminiscent of the KT/KS system
\cite{Klebanov:2000hb,Klebanov:2000nc}.  In that case, we may start in
the ultraviolet (defined with respect to a finite cutoff) with values
of the D3-brane charge $N_{UV}$ and the D5-brane charge $M$.
Following the RG flow from the UV, one finds that at intermediate
scales there is an effective D3-brane (Maxwell) charge $N_{eff}(r)$
which decreases as the theory flows to the IR.  If we assume that the
transverse space is undeformed, then at some finite radius $N_{eff}$
vanishes and the supergravity solution becomes nakedly singular.  To
avoid the singularity, the transverse conifold becomes deformed.  As a
result of the deformation, a finite scale is generated; in the field
theory, this corresponds to a dynamically generated confinement scale
which spontaneously breaks chiral symmetry.

In fact, the analogy with our example is even stronger in the case of
the cascading KS system with flavors added by inclusion of D7-branes
\cite{Ouyang:2003df,Benini:2007gx}.  There one has three types of
brane charge: D3, D5, and D7, to which one associates Maxwell charges
$N_{eff}(r)$, $M_{eff}(r)$, and $k$, respectively.  Both $N_{eff}$ and
$M_{eff}$ decrease as the theory flows to the infrared.  Now there are
two possibilities, depending on the values of $N_{UV}$ and $M_{UV}$.
If $N_{eff}$ vanishes at some radius with $M_{eff}$ finite, then the
situation is as in KS -- the naive supergravity solution is singular,
and the singularity should be resolved by deforming the conifold.  On
the other hand, it is possible for $M_{eff}$ to vanish first.  Then
the IR theory is the approximately-conformal Klebanov-Witten theory
\cite{Klebanov:1998hh} with added flavors.  This is directly analogous
to our situation, where the charges are D2, D4, and D6 in Type IIA.
In our supersymmetric solutions, $Q_2^{Maxwell,UV}$ is always large
enough that the theory flows to a superconformal fixed point in the
infrared.

This analogy leads us to conjecture that the cascading solutions dual
to three-dimensional gauge theories with $Q_2^{Page}<0$ are in fact
KT-like solutions.  Although they have naked singularities in the
infrared, it seems plausible that there exist deformed solutions which
resolve the singularity, as in KS.  The difference is that in the KS
case, it is chiral symmetry that is spontaneously broken by the
associated dynamically-generated scale, whereas in the
three-dimensional case we expect that supersymmetry will be
spontaneously broken.  (It is also possible that the field theory
exhibits a runaway behavior rather than a stable non-supersymmetric
vacuum.)

\begin{figure}
\centerline{\includegraphics{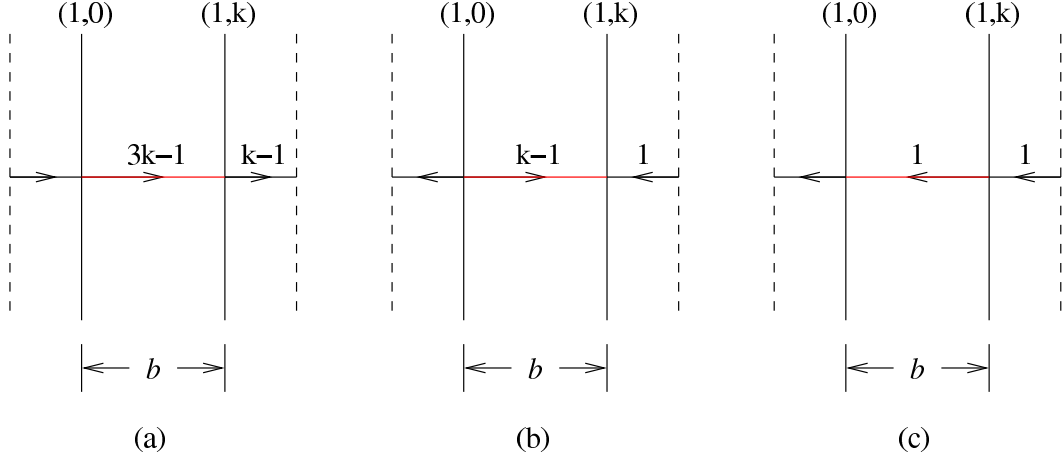}}
\caption{Hanany-Witten brane diagram for configurations violating the
generalized $s$-rule. The configurations (a), (b), and (c) are related
by sliding the $(1,k)$ brane around the circle. In this figure, labels
such as ``$3k-1$'' and ``$k-1$'' refers to the number of D3 brane
segments stretched between the 5-branes, as opposed to the counting of
integer and fractional branes. The configuration (a)
correpsonds to $N=k-1$ and $l = 2k$. Configuration (b) corresponds to $N=-1$ and $l = k$. (c) corresponds to $N=-1$ and $l=0$.
\label{figl}}
\end{figure}

One way to get a sense for the IR dynamics of these systems is to
consider the brane dynamics in the Hanany-Witten construction at the
bottom of the cascade when SUSY is expected to be broken. Consider,
for example, the configuration illustrated in figure
\ref{figl}.a. Since $N=k-1$ and $l = 2k$, the inequality $-1 = N -
l(l-k)/2k > 0$ is violated. It is easy to see that this system is
related by slides and shifts (using the terminology of
\cite{Aharony:2009fc}) to configurations illustrated in figures
\ref{figl}.b and \ref{figl}.c where the orientation of branes is such
that one expects supersymmetry to be broken. It is not immediately
clear which of the three configurations illustrated in figure
\ref{figl} is the last step of the duality cascade. Nonetheless, one
can expect some general features, i.e. the presence or absence of mass
gaps, to be shared.

One thing we infer from looking at the configuration illustrated in
figures \ref{figl}.c is that it is of the ``Borromean'' type in the
nomenclature of \cite{Mukhi:2000te}. In the absence of anti-D3-branes,
the configuration is supersymmetric, and can be described in terms of
LWY, $TN \times TN$, or $A_8$ geometry in type IIA, as we described in
the earlier section of this article.  In the IIA language, the
anti-D3-brane becomes an anti-D2. It is difficult to account for the
gravitational back reaction of the anti-D2-brane, except in one
special case. That arises when $b_\infty$ is tuned so that $m=0$,
i.e. the self-dual 4-form is tuned to zero. Then, even when the
relative sign between the warp factor and the 4-form sourced by the
2-brane is opposite of what is required for preserving SUSY, the
equation of motion for the warp factor decouples from the other fields
and takes on a simple harmonic form. The resulting geometry is such
that it is skew-whiffed $AdS_4$ in the near core region
\cite{Berkooz:1998qp,Berkooz:1999ji}. This is consistent with the
physical picture illustrated in figure 1 of \cite{Berkooz:1999ji}
where non-BPS conformal fixed points are reached through fine tuning
of some parameter which in our case turns out to correspond to
$b_\infty$, related to the relative magnitude of the gauge coupling
for the product gauge group.

\begin{figure}
\centerline{\includegraphics{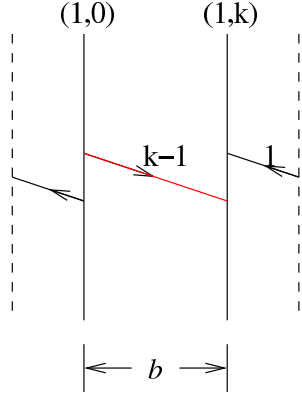}}
\caption{Schematic sketch of the expected minimum energy configuration for the construction illustrated in figure \ref{figl}.b including the effect of quantum repulsion between the brane segments. \label{figm}}
\end{figure}

What is really interesting, of course, is the fate the IR dynamics for
{\it generic} values of $b_\infty$ where the theory is expected not to
flow to the skew-whiffed conformal fixed point.  One can gain some
intuition by looking at figure \ref{figl}.b. Forces due to quantum
effects between D3-branes which are stretched along adjacent segments
are expected to be repulsive \cite{Giveon:1998sr}. Had the $(1,0)$ and
$(1,k)$ branes been parallel, this will cause the stretched D3-branes
to repel indefinitely in a run-away potential, but because they are at
an angle, there will also be a restoring force keeping the stretched
branes from getting too far. The resulting configuration is expected to look like what is illustrated in figure \ref{figm}.

This configuration is extremely similar to the configuration
illustrated in figure 3 of \cite{Mukhi:2000dn}.  This suggests that
the IR of this system is gapped, by the scale set by the mass of
open strings stretched between the adjacent D3-brane segments. The
Hanany-Witten brane analysis is not valid in the zero slope limit so
care is needed in applying the conclusion of such analysis to the
decoupled field theory. Nonetheless, we believe it is quite likely
that the fate of the IR for the non-BPS theory with generic
$b_\infty$ is the mass gap. One crude attempt to estimate the scale
of supersymmetry breaking is as follows. Consider the SUGRA solution
constructed by naively extrapolating the BPS solution to charges
which lie outside the parabola (\ref{eq6.1}). At least for small
\be \epsilon={k \over (l - {k\over 2} + b_\infty k)^2} \left({l(l-k)\over 2k}-N\right)\ , \ee
the position of the singularity can be estimated as being proportional
to\footnote{See footnote \ref{fn4} for the explanation for the meaning
of the dimensionful parameter $g_{YM2}^2$.}
\be \Lambda \propto g_{YM2}^2 \epsilon^{1/3} \ee
as long as the value of $b_\infty$ is not fine tuned, i.e.
\be l - {k \over 2} + b_\infty k \ne 0 \ . \ee
There is another  natural scale associated with this
repulson background, at
\be \Lambda \sim g_{YM}^2 \epsilon^{1/4} \ . \ee
At this radial scale, one can show that an anti D2-brane probe (as
well as any ordinary matter carrying no charges) will experience a
repulsive gravitational force, signalling that the gravity solution is
unphysical \cite{Behrndt:1995tr,Kallosh:1995yz,Cvetic:1995mx}.  Along
the lines of \cite{Johnson:1999qt}, such D-brane probes will stabilize
on a shell of radius $\Lambda$ with strong backreaction effects in the
interior of the shell. Because this is the scale at which the
background must receive large corrections, it is therefore quite
tempting to identify it as the dynamical scale of supersymmetry
breaking. The interesting non-trivial feature of this expression is
the scaling $\epsilon^{1/4}$.  This estimate is obtained by analyzing
the warp factor for the $A_8$ manifold (in the near core region when
$\epsilon$ is taken to be small), but the general dependence on
parameters $g_{YM2}^2$, $N$, $l$, $k$, and $b_\infty$ should hold for
the case of LWY and $TN\times TN$ as well.  It would be very
interesting to better understand behavior form both the field theory
and the dual gravity points of view \cite{nextgreatpaper2}.

Ultimately, these issues can be better resolved by finding a gravity
dual of the SUSY-breaking solutions, but this exercise is
mathematically challenging.  The problem might be tractable in a
system related to the $A_8$ manifold, as we discussed in Section
\ref{sec4}.  We hope to return to it in a future publication.

There is a strong relation between this case and the threshold
solution studied by Maldacena and Nastase \cite{Maldacena:2001pb}
(which was originally found by Chamseddine and Volkov
\cite{Chamseddine:2001hk}.)  Specifically, the claim of
\cite{Maldacena:2001pb} was that the gravity dual of $\N=1$ $U(N)$
Chern-Simons theory at level $k$ in 2+1 dimensions can be described by
a particular solution of type IIB string theory.  For generic values
of $N$ and $k$ they found that the supergravity solutions were
singular (containing some number of explicit D-branes), but that they
obtained a smooth solution with a finite-sized 3-cycle when $k=N/2$,
at the threshold of supersymmetry breaking.  This result had a natural
interpretation.  At the threshold, the Witten index is 1 and as a
result there is a unique ground state, suggesting that the theory is
confining.  On the gravity side, this was explained by the dynamical
generation of a scale corresponding to the finite-sized three-cycle,
and a mass gap associated with the scale of Kaluza-Klein modes on the
3-cycle.

We will conclude this article by listing some of the unresolved issues.
\begin{itemize}
\item Finding an analytic expression for the
self-dual 4-form on LWY with ${\cal N}=3$ supersymmetry.
\item
Finding a concrete field theoretical description for the 2+1
dimensional theory constructed by taking the decoupling limit of the
${\cal N}=4$ construction in section \ref{sec3}.
\item Identification of the microscopic Lagrangian for the
field theory dual to the near horizon limit of warped $A_8$ and
$B_8$ geometries. It would especially be useful to identify the
precise field theory interpretation of the parameter $\lambda$
for the $B_8$ theory.
\item Re-formulate the analogue of Page, Maxwell, and brane charges in the context of M-theory.
\item It would be interesting to study
the dynamics of the theory right at the threshold of supersymmetry
breaking in greater detail. It would also be useful to study the
deformations away from this threshold to linearized order along the
lines of \cite{Bena:2009xk}. This should also provide some
perspectives on enumerating the deformations of $A_8$ and $B_8$
theories in the vicinity of the SUSY breaking threshold.
\item
There are numerous generalizations of special holonomy and related
manifolds in eight dimensions, including Chamseddine-Volkov space,
Stenzel metrics, Aloff-Wallach spaces, tri-axial $spin(7)$
manifolds, gravitational soliton solutions, as well as variety of
$2+1d$ theories with various gauge and matter contents. It would be
interesting to explore the phase structure, including SUSY
breakings, of all of these models.  A useful first step in this
program is to review the quantization of Page charges.
\item
Finally, one hopes to better understand the fate of the infra-red
dynamics for the non-supersymmetric theories, corresponding to
points outside the red parabola.
\end{itemize}
It would be very
interesting to address any of these points in the near future.

\section*{Acknowledgements}

We would like to thank especially Ofer Aharony for collaboration on
related issues and for discussions at the early stage of this work.
We also thank Oren Bergman, Nick Halmagyi, Daniel Jafferis, and Oleg
Lunin for useful discussions. The work of AH and PO was supported in
part by the DOE grant DE-FG02-95ER40896. The work of SH was supported
by FNU via grant number 272-06-0434.

\appendix

\section{Smeared Green's Function on $TN \times TN$ }
\label{greensfn}

To find the warp factor for the cascading $\N=4$ supergravity
solution, we need to find the Green's function on the direct product
of two Taub-NUT spaces.  We have the important simplification that the
supergravity solution preserves the $U(1)$ isometries of the two
Taub-NUTs, so that it is sufficient to find the Green's function
smeared on the two $U(1)$ fibers.  This reduces our task to a six
dimensional problem in terms of the coordinates $\vec{w}_1,\vec{w}_2$.
The method that we use is essentially the one used in
\cite{Cherkis:2002ir} for the case of Taub-NUT $\times$ $R^4$.

The full six-dimensional Green's function satisfies
\beq
\nabla_6^2 G_6(\vec{w}_1,\vec{w}_2;\vec{w}_1',\vec{w}_2') = \delta^6(\vec{w}_i- \vec{w}_i')\ .
\eeq
For $\vec{w}_i\neq \vec{w}_i'$, $G_6$ is harmonic.  Moreover, the
Laplace operator may be written as $\nabla^2 = \nabla_{TN1}^2 +
\nabla_{TN2}^2$ which is separable, so that we can write
\beq
G_6(r,r') = \sum_p c_p A_p(w_1) B_p(w_2)
\label{gfconv}
\eeq
where
\beq
&& \nabla_{TN1}^2 A_p(w_1) = p^2 A_p\\
&& \nabla_{TN2}^2 B_p(w_2) = -p^2 B_p
\eeq
with the weighting factors $c_p$ chosen appropriately.  Note that
$A_p$ and $B_p$ satisfy the massive Laplace equation, so we first need
to find the associated massive Green's functions in ordinary Taub-NUT
space.  In particular, we need both an asymptotically decaying Green's
function, corresponding to $A_p$, and an asymptotically oscillatory
Green's function, corresponding to $B_p$.

In fact, the massive Green's function in Taub-NUT space smeared over
the $U(1)$ fiber was found long ago (for reasons completely unrelated
to ours) by Hostler and Pratt \cite{Hostler:1963zz}.  To be precise,
they considered the equation
\beq
\left(\nabla_r^2 + \frac{2q\nu}{|\vec{r}|} + q^2\right)
G(\vec{r},\vec{r}') = \delta^3(\vec{r}-\vec{r}')
\label{hpeqn}
\eeq
and found, for a particular set of boundary conditions, that
\beq
G(\vec{r},\vec{r}') =-\frac{\Gamma(1-i\nu)}{4\pi i q
|\vec{r}-\vec{r}'|} \left(-\frac{\partial}{\partial y}
+\frac{\partial}{\partial x}\right) W_{i\nu,1/2}(-iqx)
M_{i\nu,1/2}(-iqy)
\label{hpgreen}
\eeq
The variables $x$ and $y$ are defined by
\beq
&& x= r + r' + |\vec{r}-\vec{r}'|\\
&& y= r + r' - |\vec{r}-\vec{r}'| \ . 
\eeq
The $W$ and $M$ are Whittaker functions; in fact, any combination of
$W$ and $M$ solves the homogeneous PDE.

For imaginary (positive) $q$, Hostler and Pratt showed that this
form is required by various limits.  In particular, take both $x$
and $y$ to be large.  In this limit we expect the Green's function
to be decaying exponentially as a function of $\frac12 (x-y)
=|\vec{r}-\vec{r}'|$, which excludes $M_{i\nu,1/2}(iqx)$.  We can
take an alternate limit, with $x\rightarrow \infty$ but
$y\rightarrow 0$. This can be satisfied, for example, by taking
$\vec{r} \approx -\vec{r}'$ with $r, r' \rightarrow \infty.$ In this
limit, we expect the Green's function to be regular, which excludes
$W_{i\nu,1/2}(iqy)
\sim 1/y$.  Thus the appropriate decaying solution is the
Hostler-Pratt Green's function.

We are also interested in oscillating solutions (real $q$) for
constructing the Green's function in  $TN \times TN$  by convolution.
As we previously noted, any combination of the two types of
Whittaker functions $M$ and $W$ suffices to solve the equation of
motion.  We can fix the combination by requiring that at large
distance the Green's function should be the Green's function in
$R^3$ while that at short distance it should be the Green's function
in $R^4$.  The answer is
\beq
G(\vec{r},\vec{r}') =-\frac{\Gamma(1-i\nu)}{4\pi i q
|\vec{r}-\vec{r}'|} \left(-\frac{\partial}{\partial y}
+\frac{\partial}{\partial x}\right) M_{i\nu,1/2}(-iqx)
M_{i\nu,1/2}(-iqy).
\eeq

To map the notation of (\ref{hpeqn}) and (\ref{hpgreen}) to our
Taub-NUT coordinates, we should take $q=ip/R_1, \nu = ipR_1/4$ for
$A_p$ and $ q= p/(kR_2), \nu = p(kR_2)/4$ for $B_p$.

With these massive Green's functions in hand, we can follow the
convolution method to construct the massless Green's function in the
product space Taub-NUT $\times$ Taub-NUT.  Taking $\vec{r}_1$ and
$\vec{r}_2$ to be the two radial vectors in each Taub-NUT, and defining
\beq
&& x_i= w_i + w_i' + |\vec{w}_i-\vec{w}_i'|\\
&& y_i= w_i + w_i' - |\vec{w}_i-\vec{w}_i'|
\eeq
it appears that we should have
\beq
G(\vec{w}_1,\vec{w}_2;\vec{w}_1',\vec{w}_2') &=& \int dp
\frac{R_1(kR_2)\Gamma(1+{p\over 4 R_1})\Gamma(1+{i p\over 4kR_2})}{16\pi^3 ip |\vec{w}_1-\vec{w}_1'|
|\vec{w}_2-\vec{w}_2'|}\\ && \qquad
\times \left(-\frac{\partial}{\partial y_1} +\frac{\partial}{\partial
x_1}\right) M_{-{p\over 4 R_1},1/2}\left({p x_1 \over R_1}\right)
M_{-{p\over 4 R_1},1/2}\left({p y_1\over R_1}\right)\nonumber \\ &&
\qquad \times
\left(-\frac{\partial}{\partial y_2}
+\frac{\partial}{\partial x_2}\right) W_{{i p\over
4kR_2},1/2}\left(-{ip x_2\over kR_2}\right) M_{{i p\over
4kR_2},1/2}\left(-{ip y_2 \over kR_2}\right)\ . \nonumber
\eeq
By construction, $G$ is harmonic except where the primed and
unprimed coordinates coincide. The coefficients $c_p$ in
(\ref{gfconv}) were fixed by the following procedure.  In the
simultaneous limit where the primed and unprimed points coincide and
where the coincidence point is at large radius, the space $TN \times TN$
becomes $R^6$ and the massless Green's function is determined
accordingly, setting $c_p = -p/\pi$.

The form of $G$ is unfortunately rather inconvenient. It would be
pleasant if a simpler form existed, but we have not been able to
find it.

\section{Generalized  $B_{8\pm}$ Geometry}
\label{appendixb}

In this section we briefly review the construction of the $B_{8\pm}$
solutions enumerated in figure \ref{figg}. The content of this
section is a small extension of appendix A of \cite{Cvetic:2001pga}.
Consider the ansatz (\ref{a8ansatz})
\be ds^2 = h(r)^2 dr^2  + a(r)^2  (D \mu^i)^2  + b(r)^2 \sigma^2 + c(r)^2d \Omega_4 \label{eqb1}\ee
and set
\be h(r) = {\ell \over b(r)} , \qquad c(r) = \ell \sqrt{f(r)} \ .  \label{eqb2}\ee
Then, the Ricci-flatness condition reduces to\footnote{In section
\ref{sec5}, $\ell$ was denoted $\tilde \ell$ but we will drop the
tilde here. Also, note that many formulas in the literature are
reported where $\ell$, which has dimensions of length, is set to 1.}
\be 2 f^2 f''' + 2 f (f'-3\ell^{-1}) f'' - (f'+\ell^{-1})(f'-\ell^{-1})(f'-3\ell^{-1}) = 0 \label{feq}
\ee
where prime denotes derivative with respect to $r$, and $a(r)$ and
$b(r)$ can be expressed in terms of $f(r)$ as
\be a^2 = {\ell^2 (f'-\ell^{-1}) (f'-3\ell^{-1}) f \over Q}, \qquad b^2 = {2 a^2 \over \ell^2 (f'-\ell^{-1})^2} \ee
with
\be Q = 2 f W'+ (f'-3\ell^{-1}) W, \qquad W = f'-\ell^{-1} \ . \ee
The third order equation for $f(r)$ can be solved using the
following trick. Define new variables $G$ and $z$ so that\footnote{We are working in the convention where $\ell$, $r$, $a$, $b$, and $c$ have dimension of length, and $h$, $f$, $z$, $G$, and $\lambda$ are dimensionless.}
\be f(z) = z G(z)^2, \qquad r = r_0 + \ell \int_0^z G(z')^2 dz' \label{fr}\ee
Then, it can be shown that (\ref{feq}) is satisfied if $G(z)$ solves
\be
{d^2G \over dz^2} = {c \over 2}   G^3 \ . 
\ee
This is essentially a particle in a $V(G) = -G^4$ potential, which can
be brought to a first order form using conservation of energy.
\be {dG \over dz} = \sqrt{2 E + {c \over 4} G^4}  \ . \ee
Without loss of generality or affecting (\ref{fr}), $E$ can be set
to $\pm 1$ by rescaling
\be G \rightarrow |E|^{1/6} G, \qquad z \rightarrow |E|^{-1/3} z, \qquad c \rightarrow |E|^{1/3} \lambda \ . \ee
Let us first take $E>0$. so that
\be {dG \over dz} = \sqrt{2  + {\lambda \over 4} G^4}  \ee
Then, we can solve for $z(G)$ by
\be z(G)  = z_0 + \int_0^G dG' \, {1 \over \sqrt{2 + {\lambda \over 4} G'^4}}  = z_0 + {G \over \sqrt{2}} {}_2 F_1(\tfrac{1}{4},\tfrac{1}{2},\tfrac{5}{4}, -{\lambda \over 8} G^4 ) \label{elliptic}
\ee
which can be inverted, by recognizing the fact that
(\ref{elliptic}) can be interpreted as an elliptic integral
\cite{abramowitz+stegun}, to an expression of the form
\be G(z) = -2^{3/4} \lambda^{-1/4} {1 - \mbox{tn}(\sqrt{2} \lambda^{1/4} k^{-1}(z- z_0+\zeta),k) \over
1 + \mbox{tn}(\sqrt{2} \lambda^{1/4} k^{-1}(z - z_0+\zeta) ,k)}
\ee
where
\be k = {2^{5 /4} \over \sqrt{2}+1}, \qquad
\zeta = \lambda^{-1/4} {\Gamma({1 \over 4})^2 \over 2^{11/4} \sqrt{\pi}}\ . \ee

The undetermined constants in solving the third order equation
(\ref{feq}) are accounted for by the integration constants $r_0$,
$z_0$, and $\lambda$. Of the three, the shift of $r_0$ leads only to
change of coordinates and does not change the geometry. The remaining
two, $z_0$ and $\lambda$ parametrizes the sizes of $S^1$ at infinity
and $S^4$ at the core. The size of $S^1$ at infinity can be inferred
from the behavior of $b(r)$ from which we infer
\be L^{-2} = b(r=\infty)^{-1} = {\lambda \over 2}  (z_0 + \zeta) \ell^{-2} \ .
\ee
One can eliminate $z_0$ in terms of $L$, and so we have $L$ and
$\lambda$ parameterizing this family of solutions. Keeping $L$ fixed
(to $\ell$, without loss of generality since rescaling of $\ell$ can
be absorbed into reparameterization of $r$) and varying $\lambda$ is
what is illustrated in figure \ref{figg}. In this parametrization,
$B_{8\inf}$ corresponds to taking
\be \lambda = \lambda_\infty \equiv {8 (2 \pi)^{2/3} \over \Gamma({1\over 4})^{8/3}} \ee
whereas $\lambda \rightarrow \infty$ limit corresponds to $B_8$. To
find the $B_{8-}$ solutions, we take the case where $E$ is negative and
\be G(z)  ={2^{3/4} \lambda^{-1/4} \over {\rm cn} (2^{1/4} \lambda^{1/4} (z-z_0+\zeta),2^{-1/2}) }
\ee
with $\lambda$ ranging from $0$ to $\infty$.

\begin{figure}
\centerline{\includegraphics[width=2.7in]{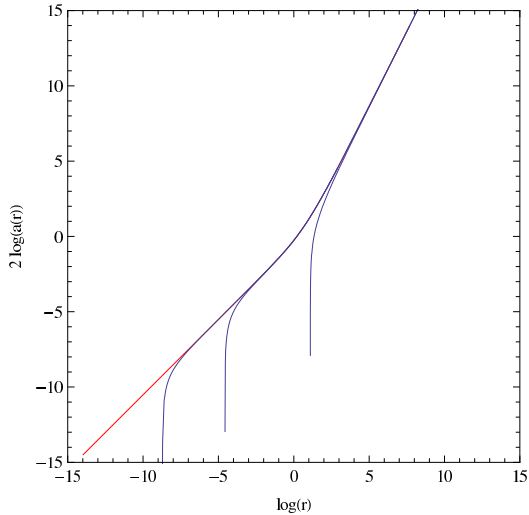}}
\caption{Log-log plot of $a(r)$ for $B_{8 \pm\infty}$. The red curve corresponds to $B_{8\infty}$ and describes the cross-over from the $a(r)^2 \sim r^2/4$ behavior at large $r$ and the squashed $S^7$ cone behavior for small $r$.   The other $B_{8\pm}$ solutions can be viewed as a deformation of $B_{8\infty}$ in the IR. \label{figi}}
\end{figure}

Using these expressions, one can determine $a$, $b$, $c$, and $h$
analytically, using either $z$ or $G$ to parametrize the radial
coordinate. After determining $a$, $b$, $c$, $h$, and $r$ in terms of
$z$ or $G$, it is straight forward to show that all of these
geometries have the same large distance asymptotics as the $B_8$, i.e.
\be a(r)^2 \sim {r^2 \over 4}, \qquad b(r)^2 \sim \ell^2, \qquad c(r)^2 \sim {1 \over 2} r^2 \ . \ee
The core of these geometries are defined by the point where $a(r)=b(r)=0$ and this will depend on the value of $\lambda$.  A plot of $a(r)$ in log-log scale is illustrated in figure \ref{figi}. The core can be inferred form the point where $a(r)$ approaches zero rapidly in the plot.

These geometries have two scales, the asymptotic size of $S^1$ and the
size of $S^4$ at the core.  In \cite{Cvetic:2001pga,Cvetic:2001zx},
emphasis was placed on deforming among $B_{8\pm}$ keeping the size of
the $S^4$ fixed, which meant that the limiting case of $B_{8+}$
amounted to taking the size of $S^1$ to infinity, changing the
asymptotic geometry. Once can alternatively consider sending the size
of $S^4$ to zero keeping the size of $S^1$ fixed. This latter scaling
gives rise to the $B_\infty$ geometry. These two limits are the same
in that the ratio of the sizes of $S^4$ to $S^1$ are going to zero,
and differ only in the scaling of the radial coordinate.

The last detail about the $B_{8\pm\infty}$ that is relevant to our
discussion is the construction of self-dual and anti-self-dual 4-forms
and their implications for the repulson-free condition.  Since the
explicit metric of the $B_{8\pm\infty}$ background are available, one
should be able to find these 4-forms by imposing (anti)self-duality
condition on the 4-form $dC$ where $C$ is given in
(\ref{3formansatz}).  In practice, this procedure is somewhat
cumbersome.   One can write the 4-form as
\beq dC &=& m \left[ u_1 (h a^2 b\,  dr \wedge \sigma \wedge X_2 \pm c^4 \, \Omega_4) + u_2 ( h b c^2 \, dr \wedge \sigma \wedge Y_2 \pm a^2 c^2 \, X_2 \wedge Y_2) \right. \cr
&& \left. +u_3 (h a c^2 \,dr \wedge Y_3 \mp a b c^2 \, \sigma \wedge X_3)\right] \label{b18}\eeq
and in  terms of variables
\be U_1 = u_1 c^4, \qquad U_2 = u_2 a^2 c^2, \qquad U_3 = u_3 a b c^2\ee
the (anti)-self-duality condition takes the form
\be U_1 = \pm (4 v_3 - 2 v_2), \qquad U_2 = \pm(v_2 - v_1 + 2 v_3), \qquad U_3 =\pm(v_1+v_2)\ee
and
\be {d \over dr} U_i = M_{ij}(r) U_j \label{eqMr}\ee
where
\be
M = \pm \left(\begin{array}{ccc} 0 & 2m_1 & 4 m_3 \\ - m_1 & m_2 & 2 m_3 \\ m_1 & m_2 & 0 \end{array}\right) , \qquad m_1 = {a^2 \over f^2}, \qquad m_2 = {1 \over a^2} , \qquad m_3 = {1 \over b^2}
\ee
and
$a$, $b$, $f$, etc are as given in (\ref{eqb1}) and
(\ref{eqb2}). Here, $+$ and $-$ refers to self-dual and
anti-self-dual, respectively. Since (\ref{eqMr}) has the form of the
time dependent Schrodinger equation, it can be solved formally in
terms of the Dyson's path ordered exponential. Alternatively, one can
resort to analyzing this equation numerically (as we show below).

Fortunately, one can infer enough information about the repulson-free
condition from general consideration alone.

\begin{figure}
\centerline{\begin{tabular}{cc}
\includegraphics[width=3in]{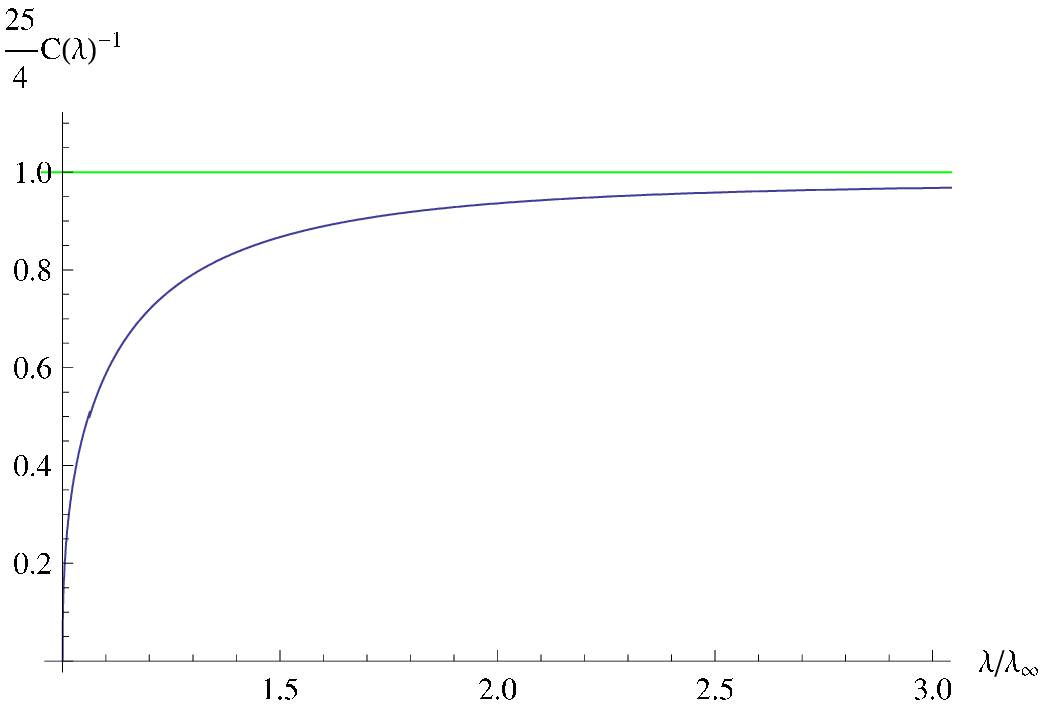} &
\includegraphics[width=3in]{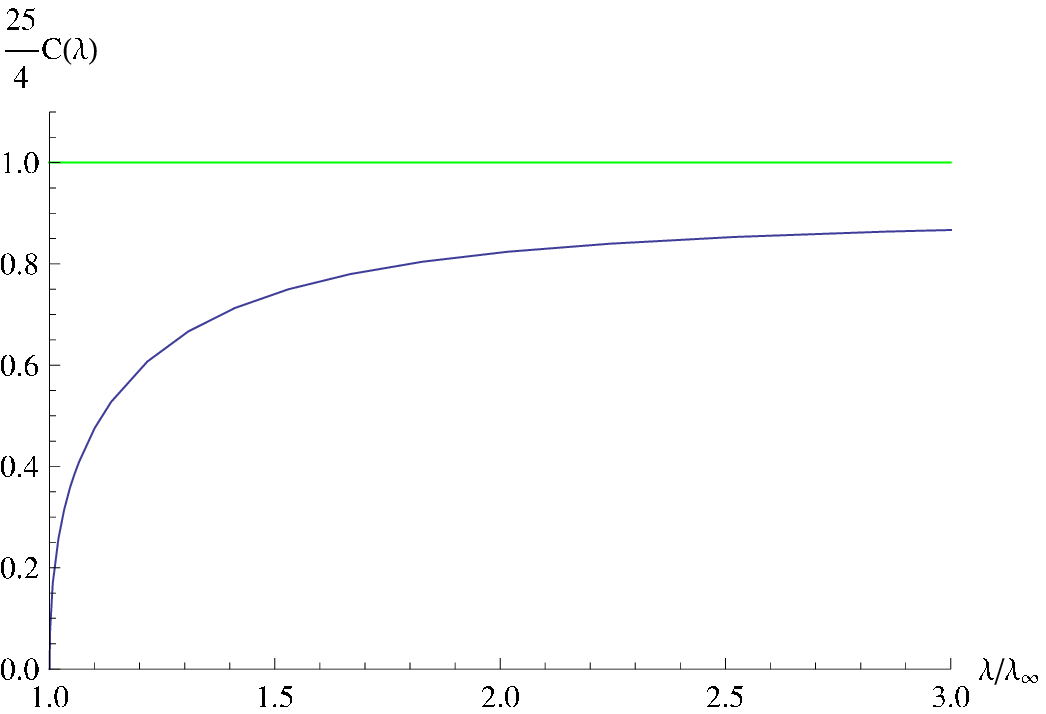} \\
(a) & (b)
\end{tabular}}
\caption{$C(\lambda)$ evaluated numerically for the (a) anti-self-dual and (b) self-dual 4-forms on a family of deformed $B_8$ space  parameterized by $\lambda$. $\lambda = \lambda_\infty$ corresponds to the $B_{8\infty}$ limit and $\lambda=\infty$ corresponds to the $B_8$ limit. The analysis shows that $C(\lambda)$ are asymptoting toward the expected values of $25/4$ and $4/25$, respectively, for the anti-self-dual and self-dual 4-forms. In the $B_{8\infty}$ limit, $C(\lambda)$ is going to $\infty$ and $0$, respectively, for the anti-self-dual and self-dual 4-forms. The divergence of $C(\lambda)$ for the anti-self-dual 4-form in the $B_{8\infty}$ limit is a reflection of the fact that the anti-self-dual 4-form is becoming non-normalizable. \label{figj}}
\end{figure}

The repulson condition concerns the D2 flux at the core. The flux at
infinity minus the flux at the core is the total bulk contribution to
D2 charges coming from the $G_4 \wedge G_4$ term.  We expect on
general grounds that the D2 Maxwell charge at infinity is proportional
to
\be N+{k \over 8}  + b_\infty\left (l - {k \over 2}\right)
+ {1 \over 2} b_{\infty}^2 k = N - {l (l-k) \over 2k}
+ {1 \over 2k} \left(l - {k \over 2} + b_\infty\right)^2\ . \ee
The contribution from the integral of $G_4 \wedge G_4$ is proportional
to $m^2$ and so the flux at the origin should take the form
\be  N - {l (l-k) \over 2k} + {C(\lambda) \over 2k} \left(l - {k \over 2} + b_\infty\right)^2 \ .  \label{fluxcore}\ee
This is what was anticipated in (\ref{repulsonfree}).  It can be shown
that the value of $C(\lambda)$ is related to the value of $v_1(r)$ at
the core\footnote{This can also be viewed as arising from the 2-brane
charge induced on world volume of D4 and D6 brane by the value of the
$B_{NSNS}$ at the core through the Wess-Zumino term.}
\be C(\lambda) = \left. \rule{0ex}{3ex} 64 v_1(r)^2 \right|_{r={\rm core}} \ee
e.g. for anti-self-dual case, provided that the $v_i(r)$ are tuned to
asymptote in the large $r$ region to match the anti-self-dual 4-form
(\ref{b8asd}) and that the 4-forms be normalizable. Similar relation
holds for the self dual case.  This quantity can be computed by
numerically solving (\ref{eqMr}).\footnote{For the anti-self-dual
case, the analysis can be simplified drastically by exploiting the
identity $u_1+2 u_2- 4 u_3=0$ \cite{Cvetic:2001pga}.}  The result is
illustrated in figure \ref{figj}. In this parametrization, $B_{8}$
limit corresponds to sending $\lambda \rightarrow \infty$. The result
of our numerical analysis show that $C(\lambda)$ is asymptoting to the
expected value in the $B_8$ limit.

Independent of the precise value of $C(\lambda)$, one can infer from
the form of (\ref{fluxcore}) that the parabola describing the
repulson-free region for the self-dual and anti-self-dual 4-forms will
touch at one, and only one, point, as is illustrated in \ref{figh},
for generic values of $\lambda$.

There are several general lessons one can infer from the form of
(\ref{fluxcore}), the existence of one parameter family of deformation
of $B_8$ space illustrated in figure \ref{figg}, the general layout of
the parabolas as is illustrated in figure \ref{figh}, and the
numerical values of $C(\lambda)$ as is illustrated in figure
\ref{figj}. These lessons can be summarized as follows

\begin{itemize}
\item Unlike in the case of the $A_8$, the position of the parabola will depend on the values of $b_\infty$ because $C(\lambda)$ is non-vanishing in general.
\item In the $\lambda \rightarrow \lambda_\infty$ limit where $B_8$ asymptotes to $B_{8\infty}$,  the red parabolas will degenerate and disappear from the phase diagram since $C(\lambda)$ is diverging there.
\item Also, in the $B_{8\infty}$ limit, the position of the parabola is independent of $b_\infty$ since $C(\lambda)$ is going to zero there.
\end{itemize}

The region inside the green parabola in the $B_{8\infty}$ limit
appears to be special, where the geometry asymptotes in the IR to the
warped squashed cone of \cite{Ooguri:2008dk}, except that the D2
charge are such that the geometry is skew-whiffed.  In other words,
although the IR of the $B_{8 \infty}$ theory appears to be breaking
supersymmetry, it appears nonetheless to be conformal.

\begin{figure}[t]
\centerline{\includegraphics[width=3in]{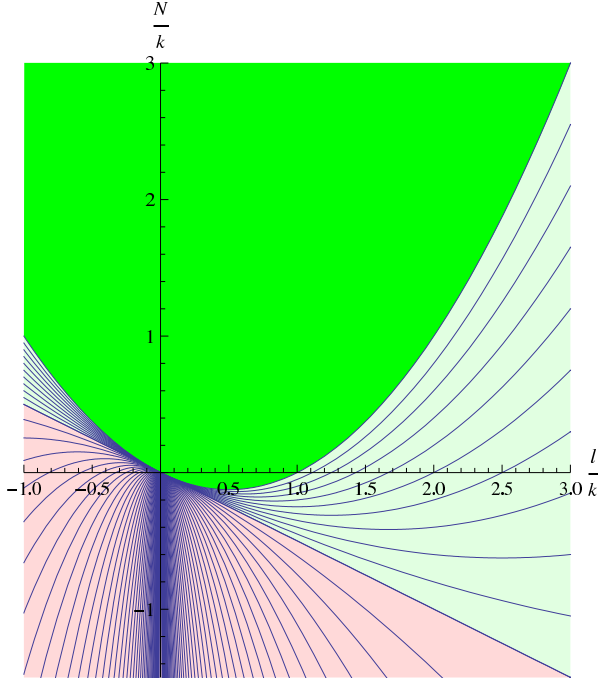}}
\caption{By tuning $\lambda$ for fixed $N$,$l$, $k$, and
    $b_\infty$, one can arrange to pick a point in the phase diagram
    corresponding to the edge of the parabolic region. The regions
    illustrated in light green are the boundaries of green parabolic
    region, and the regions illustrated in light red are the
    boundaries of the red parabolic region. By setting
    $\lambda=\lambda_\infty$ in the green parabolic region, $\lambda$
    as a function of $x=N/k$, $y=l/k$, and $b_\infty$ will be
    continuous but non-analytic.  The collection of blue parabolic
    lines corresponds to the contour of fixed $\lambda$. All possible
    range of $\lambda$'s including both $B_{8+}$ and $B_{8-}$, are
    reflected in this plot.
\label{figk}}
\end{figure}

The broad picture is that for every choice of $N$, $l$, $k$, $q$, and
$b_\infty$, there is specific dynamical system for which we have
partial understanding of the dynamics through the dual supergravity
description. $N$, $l$, $k$, and $q$ are discrete
parameters. Continuous parameters $b_\infty$ and $\lambda$ are
constrained once all the discrete parameters are fixed.  Alternatively
for fixed $k$, $b_\infty$, and $\lambda$, which constrains $q$, one
can illustrate the phase diagram as a function of $N$ and $l$ as is
illustrated in figure \ref{figh}. As we noted in number of contexts,
the points along the boundary of the parabolic region are special from
the gravity point of view, in that the 2-brane source at the core is
identically zero. Remarkably, it turns out that one can chose a unique
value of $\lambda$ for (almost) every choice of $N$, $l$, $k$, and
$b_\infty$ so that we sit at this special point on the boundary of the
parabolic region as is illustrated in figure \ref{figk}. The regions
illustrated in light red and light green corresponds to various slices
of the boundary of the red and green parabolas of figure
\ref{figh}. This foliation turns out not to cover the entire range of
$N$ of $l$ because as $\lambda$ approach $\lambda_\infty$, the green
parabola for the repulson-free region of the $B_{8\infty}$ do not
collapse to zero size. However, a natural extension of $\lambda$ as a
function of $N$, $l$, $k$, and $b_\infty$ is to set it equal to the
constant value $\lambda_\infty$. In this way, $\lambda$ as a function
of these variables will be continuous, giving rise to a smoothly
connected family of supergravity solutions. The supergravity solution
itself will also be singularity free, except at the boundary of the
$B_{8\infty}$ parabola where there will be a conical singularity,
whose base is the squashed $S^7$, in the core region.

For all of the points parameterized by $N$, $l$, $k$, $b_\infty$, and
$\lambda$ in the phase diagram illusterated in figure \ref{figh}, the
geometry in UV asymptotes to that of $B_8$ which preserves ${\cal
N}=1$ $d=3$ supersymmetry. For all points outside the red parabolas,
however, this supersymmetry is broken. Since in the perspective of the
field theory dual, this is dynamical breaking of supersymmetry, one
expects to find the corresponding Goldstone fermions. At least in the
region inside the green parabola of figure \ref{figh}, a gravity
description of these theories in the broken supersymmetry phase is
availble. For these geometires, one can show that there are precisely
two normalizable fermion zero modes, following the analysis of
\cite{Duff:1995wk}.

In \cite{Cvetic:2001zx}, a ``tri-axial'' family of $spin(7)$ holonomy
manifold was constructed, further broading the arsenal of known
$spin(7)$ holonomy manifolds.  These authors then identified a one
parameter subset of this family, which they named $C_8$. We expect
much of what we found for the $B_{8\pm\infty}$ to carry over to the
$C_8$ case as well.

\bibliography{cascade}\bibliographystyle{utphys}

\end{document}